\documentclass[a4paper,11pt,preprint]{article}
\usepackage{jcappub}
\usepackage[toc,page]{appendix}
\usepackage{aas_macros}

\usepackage[font=scriptsize]{caption}
\usepackage{dsfont}
\usepackage{graphicx}
\usepackage{dcolumn}
\usepackage{bm}
\usepackage{color}
\usepackage{epsfig}
\usepackage{tablefootnote}
\usepackage[colorlinks = true,
            linkcolor = blue,
            urlcolor  = blue,
            citecolor = blue,
            anchorcolor = blue]{hyperref}
\usepackage{natbib}
\usepackage{float}
\usepackage{footmisc}
\usepackage[utf8]{inputenc}
\usepackage{tikz-feynman}
\usepackage{comment}
\usepackage[normalem]{ulem}

\definecolor{lightred}{rgb}{1,0.5,0.5}
\definecolor{lightgreen}{rgb}{0.5,1,0.5}
\definecolor{lightblue}{rgb}{0.5,0.5,1}
\definecolor{lightcyan}{rgb}{0.5,0.75,0.75}
\definecolor{lightmagenta}{rgb}{0.75,0.5,0.75}
\definecolor{customgreen}{rgb}{0.494,1,0.502}

\newcommand{\meV}{\mathinner{\mathrm{meV}}}
\newcommand{\eV}{\mathinner{\mathrm{eV}}}
\newcommand{\keV}{\mathinner{\mathrm{keV}}}

\newcommand{\GeV}{\mathinner{\mathrm{GeV}}}

\usepackage{todonotes} 

\usepackage{comment}
\usepackage{hyperref}

\title{Spectra of axions emitted from\\ main sequence stars}

\author{Ngan H. Nguyen\footnote{email: \href{mailto:nnguye53@jhu.edu}{nnguye53@jhu.edu}}, Erwin H. Tanin\footnote{email: \href{mailto:ehtanin@gmail.com}{ehtanin@gmail.com}}, and Marc Kamionkowski}
\affiliation{William H. Miller III Department of Physics and Astronomy,\\
Johns Hopkins University, Baltimore, MD 21218, USA.
}

\abstract{
We compute the detailed energy spectra of axions with two-photon coupling produced in stellar cores over a wide range of stellar masses. We focus on main sequence stars and base our calculations on the stellar interior profiles from MESA, for which we provide simple fits in an appendix. The obtained stellar axion spectra, combined with recent models of star formation history and stellar initial mass function, enable us to estimate the properties of the diffuse axion background sourced by all the stars in the universe. The fluxes of this stellar axion background and its decay photons are subdominant to but can in principle be disentangled from those expected from the Sun and the early universe based on their different spectral and spatial profiles.
}
\begin{document}

\maketitle

\newpage

\section{Introduction}
New light particles with feeble interactions arise ubiquitously in a wide range of beyond the Standard Model theories that address various issues of the Standard Model \cite{Essig:2013lka,Jaeckel:2010ni}. To maximize the discovery potential of these particles, it is useful to have a quantitative understanding of their production from all possible sources across all energy regimes. Stars are among the most intense continuous sources of new light particles in the present epoch \cite{Raffelt:1996, AxionLimits}. While the impact of new particle emission on stellar evolution has been extensively explored, a characterization of the energy spectra of the particles emitted from stars other than the Sun has been lacking \cite{CAST:2007jps, Hoof:2021mld, Hoof:2023jol, OHare:2020wum, Caputo:2020quz, Bastero-Gil:2021oky, DiLella:2000dn,  Lopes:2021mgy, Schwarz:2015lqa, Redondo:2015iea}. Knowing the properties of such star-sourced light particles can be useful as it may reveal new opportunities for probing beyond the Standard Model.

The aim of this work is to provide benchmark stellar-emission spectra of light particles from main sequence stars over a wide range of stellar masses. We focus, as a start, on what we refer to as the \textit{axion}, a pseudoscalar $a$ whose Lagrangian includes \cite{Arvanitaki:2009fg,Bauer:2004tk,Brivio:2017ije,Agrawal:2017cmd,Agrawal:2022lsp}
\begin{align}
    \mathcal{L}\supset \frac{1}{2}\partial_\mu a\partial^\mu a-\frac{1}{2}m_a^2a^2-\frac{g_{a\gamma\gamma}}{4}aF_{\mu\nu}\tilde{F}^{\mu\nu},
\end{align}
where $F_{\mu\nu}$ is the electromagnetic field strength tensor and the axion-photon coupling $g_{a\gamma\gamma}$ is treated as independent from the axion mass $m_a$.\footnote{All the results we obtain below apply as well to a CP-even scalar $s$ with purely electromagnetic coupling of the form $g_{s\gamma\gamma}s F_{\mu\nu}F^{\mu\nu}$, since this coupling leads to processes analogous the axion processes that we consider with the same amplitudes \cite{Marsh:2014gca,Brax:2010xq,Caputo:2021rux}. This is, of course, assuming the scalar bare mass is greater than the radiative corrections it may receive.} CAST \cite{CAST:2007jps,CAST:2017uph} and globular cluster observations \cite{Ayala:2014pea,Dolan:2022kul} have essentially ruled out $g_{a\gamma\gamma}\gtrsim 6.6\times 10^{-11}\GeV^{-1}$ for $m_a\lesssim 10\keV$, nevertheless the remaining parameter space still allows axions to be produced at some level in stellar cores and leave interesting astrophysical signatures \cite{AxionLimits, Raffelt:1996}. 

Over the past decades, our quantitative understanding of stellar structure and evolution has improved dramatically, largely thanks to important calibrations of stellar evolution models against asteroseismic data enabled by the recent advent of space-based photometry \cite{2021RvMP...93a5001A, 2003PASP..115.1023W,2009A&A...506..411A,2010Sci...327..977B,2014PASP..126..398H,2015JATIS...1a4003R}. The majority of asteroseismic studies have been on main sequence stars. Post-main sequence, nuclear-burning stars are statistically rare in numbers due to their short lifetimes, and are generally less understood \cite{Nugis_2002,2011A&A...526A.156M,Yoon:2017dme,McDonald_2018,Hofner:2018,Beasor:2022sey,Prager_2022}. One major reason for the latter is that the underlying physical mechanism behind the strong mass loss that is known to occur for these stars is not yet understood.\footnote{Most stellar evolution codes treat mass loss very crudely, controlled by free parameters that are yet to be empirically calibrated. Moreover, there are significant discrepancies among the mass loss rates predicted by different codes \cite{2017A&A...603A.118R}.} For these reasons, we restrict our analysis to main sequence stars. While mass loss also plays a substantial role for high-mass main sequence stars, these stars are better constrained due to the better availability of their asteroseismic data \cite{2020FrASS...7...70B}. The key input parameters of stellar evolution models include the initial mass, metallicity, rotation, and magnetic field. To simplify our analysis, we consider only the dominant parameter of these, namely the initial stellar mass, neglect the rotation and magnetic field, and set the metallicity to that of the cosmic average.

Stars, collectively, can also be regarded as a cosmic source of axions.\footnote{Cosmic background of star-sourced particles have been studied previously for neutrinos \cite{Brocato:1997tu, Porciani:2003zq, Iocco:2004wd,Iocco:2007td,Nakazato:2005ek}. } The resulting \textit{Stellar Axion Background} (StAB) spectrum is a triple integral over the interior of a star, the stellar population at a particular epoch, and the star formation history. These are characterized by a stellar-evolution model, a stellar initial mass function, and a star formation rate. The StAB will contribute to the diffuse extragalactic axion background together with other potential cosmic axion sources, which include supernovae \cite{Raffelt:2011ft, Calore:2020tjw,Calore:2021hhn,Kolb:1988pe,Lella:2022uwi,DeRocco:2019njg,Diamond:2023scc}, dark matter \cite{ Nomura:2008ru,Fitzpatrick:2023xks,Ghosh:2023tyz,Levkov:2016rkk, Fox:2023aat,Buch:2020mrg}, dark energy fluctuations \cite{Berghaus:2020ekh,Ji:2021mvg}, primordial black holes \cite{Agashe:2022phd,Jho:2022wxd}, and various other astrophysical \cite{Caputo:2021kcv,Diamond:2023cto} and early universe \cite{Balazs:2022tjl,Langhoff:2022bij, Dror:2021nyr,Cadamuro:2011fd,Angus:2013sua, Conlon:2013txa,Jaeckel:2021ert} processes. One can in principle distinguish the StAB from other axion backgrounds based on their spectra and spatial distributions. If the axion is sufficiently heavy, a considerable fraction of the StAB can spontaneously decay into X-ray within the age of the universe. These X-ray photons will contribute to the cosmic X-ray background (CXB) and potentially leave an imprint in the form of a local bump in the CXB spectrum.

The paper is organized as follows. We calculate the axion spectra from main sequence stars over a wide range of stellar masses individually and collectively in Section~\ref{s:axionspectra}, evaluate the detectability of the X-ray from the decay of stellar axions in Section~\ref{s:StABXray}, and conclude in Section~\ref{s:conclusion}. Fits to the interior profiles of the ensemble of main sequence stars used in our analysis are collected in Appendix.~\ref{AppendixA}.

\section{Axions from main sequence stars}
\label{s:axionspectra}

\subsection{Axion production in stars with different masses}

Axions are produced in stellar cores primarily through the Primakoff process (thermal photons converting into axions in the static electric fields sourced by charged particles in the star) and photon coalescence $\gamma\gamma\rightarrow a$. The axion production rate per unit volume via the Primakoff effect\footnote{Our energy-integrated axion emission rate from the Primakoff process is in agreement with the total axion luminosity of \cite{CAST:2007jps} as well as the axion luminosity per unit stellar interior mass of \cite{DiLuzio:2021ysg} which, as pointed out in \cite{2017A&A...605A.106C}, is about an order of magnitude larger than the expression reported in \cite{Friedland:2012hj,2015PhRvD..92f3016A}.} \cite{Cadamuro:2011fd, Jaeckel:2017tud, Carenza:2020zil,Dolan:2021rya} and photon coalescence\footnote{Until recently \cite{Bastero-Gil:2021oky,  Caputo2022,Muller2023, Ferreira2022}, axion production in dense astrophysical objects via photon coalescence has mostly been neglected, as this process is negligible by far compared to axion production via the Primakoff effect when the axion is light. However, for heavier axions with masses comparable to or higher than the core temperatures of stars, photon coalescence process can be more efficient than the Primakoff effect.} \cite{Bastero-Gil:2021oky,Caputo2022,Muller2023,Ferreira2022} as functions of the axion energy $E_a$ are well known and can be written as
\begin{align}
    \frac{d\dot{n}_a^{\rm Prim.}}{dE_a}=\ &\frac{g_{a\gamma\gamma}^2\kappa^2TE_a \sqrt{E_a^2-\omega_p^2}}{32\pi^3\left(e^{E_a/T}-1\right)}\left[\left(1-\frac{1}{2\gamma_a^2}+\frac{\kappa^2}{4\gamma_a^2m_a^2}\right)\ln\left(\frac{2\gamma_a^2(1+v_a)-1+\kappa^2/m_a^2}{2\gamma_a^2(1-v_a)-1+\kappa^2/m_a^2}\right)\right.\nonumber\\
    &\left.-\frac{m_a^2}{4\gamma_a^2\kappa^2}\ln\left(\frac{2\gamma_a^2(1+v_a)-1+m_a^2/\kappa^2}{2\gamma_a^2(1-v_a)-1+m_a^2/\kappa^2}\right)-v_a\right],\label{Primakoff}\\
    \frac{d\dot{n}_a^{\rm coal.}}{dE_a}=\ &\Theta\left(m_a-2\omega_{\rm p}\right)\frac{g_{a\gamma\gamma}^2 T m_a^2(m_a^2-4\omega_{\rm p}^2)}{64\pi^3(e^{E_a/T}-1)}\ln\left[\frac{\text{sinh}\left[\gamma_a( m_a+ v_a\sqrt{m_a^2-4\omega_{\rm p}^2})/4T\right]}{\text{sinh}\left[\gamma_a(m_a-v_a\sqrt{m_a^2-4\omega_{\rm p}^2})/4T\right]}\right],\label{Coalescence}
\end{align}
where $v_a=\sqrt{1-1/\gamma_a^2}$ and $\gamma_a=E_a/m_a$ are respectively the velocity and the corresponding Lorentz factor of the axion; $\kappa$ and $\omega_{\rm p}$ are respectively the inverse screening length and the plasma mass in the stellar interior, given by
\begin{align}
    \kappa^2=4\pi\alpha\frac{\sum_{i=e,\text{ ions}} Z_i^2 n_i}{T},\quad\quad \omega_{\rm p}^2=4\pi\alpha\frac{n_e}{m_e},
\end{align}
where $T$, $Z_i$, and $n_i$ are the temperature, charge of species $i$ (in units of electron charge), and number density of species $i$. The spectral axion emission rate from a whole star is then found by integrating over the volume of the star
\begin{align}
    \frac{d\dot{N}_a^{\star}}{dE_a}=\int_{\rm star} dV\; \left(\frac{d\dot{n}_a^{\rm Prim.}}{dE_a}+\frac{d\dot{n}_a^{\rm coal.}}{dE_a}\right),\label{Nasingle}
\end{align}
which requires knowing the internal profiles of the star.

We obtain the stellar interior profiles from the state of the art stellar evolution code Modules for Experiments in Stellar Astrophysics (MESA) \cite{Paxton2011,Paxton2013, Paxton2015, Paxton2018, Paxton2019, Jermyn2023}.\footnote{We use MESA r22.11.1 version in this paper.}  MESA is a one-dimensional, i.e. spherically symmetric, stellar evolution code which numerically solves the coupled equations for the structure, nuclear reaction network, and energy transfers (convection, radiative transfer, mass loss) of individual stars. We generate the profiles of 34 representative main-sequence stars with masses ranging from $0.1-100 M_\odot$ using MESA. The inputs of the simulation are chosen so as to produce the most typical main-sequence stars.\footnote{We adopt the following assumptions in running the MESA code. The initial metallicity of the stars are uniformly set to the cosmic average $\langle Z\rangle=0.0175$ taken from \cite{Calura:2004jc}. The helium abundance is set to MESA's default $Y= 0.24 + 2Z$. The radiative opacities are taken from the standard Type 1 OPAL opacity tables \cite{1996ApJ...464..943I} based on the solar chemical compositions from \cite{Asplund_2009}, with the setting such that it automatically switches to Type 2 OPAL \cite{1993ApJ...412..752I} when appropriate.} We simulate the evolution of these stars starting from their slowly-contracting, pre-main-sequence stage. At some point during the evolution, hydrogen burning ignites and halts the contraction, marking the start of the main sequence phase. We let the stars evolve through the entire main sequence phase and define the end of the phase at the age $t_{\rm life}$ when the central hydrogen fraction reaches $X=10^{-4}$.

Axion production in a star depends mainly on the temperature $T(r)$, screening length $\kappa(r)$, and plasma mass $\omega_{\rm p}(r)$ profiles in the core region of the star. Our simulations show that these quantities evolve slowly by $O(10\%)$ over the bulk of the main sequence stage and only start to vary appreciably toward the end of the stage. To reduce the computational cost of evaluating the stellar axion production, we neglect this time dependence and extract the stellar profiles from a representative point in the stellar evolution. Whenever possible, the profiles of these stars are taken from a snapshot at the so-called intermediate age main sequence phase, namely the point when the hydrogen abundance hits $X=0.3$. For low-mass stars $M<M_\odot $ which do not reach this phase within the age of the universe due to their slow evolution, the profiles are instead extracted at half the age of the universe, $t_\text{U}/2=6.85 \text{ Gyr}$.

\begin{figure}
   \centering
  \includegraphics[width=0.95\linewidth]{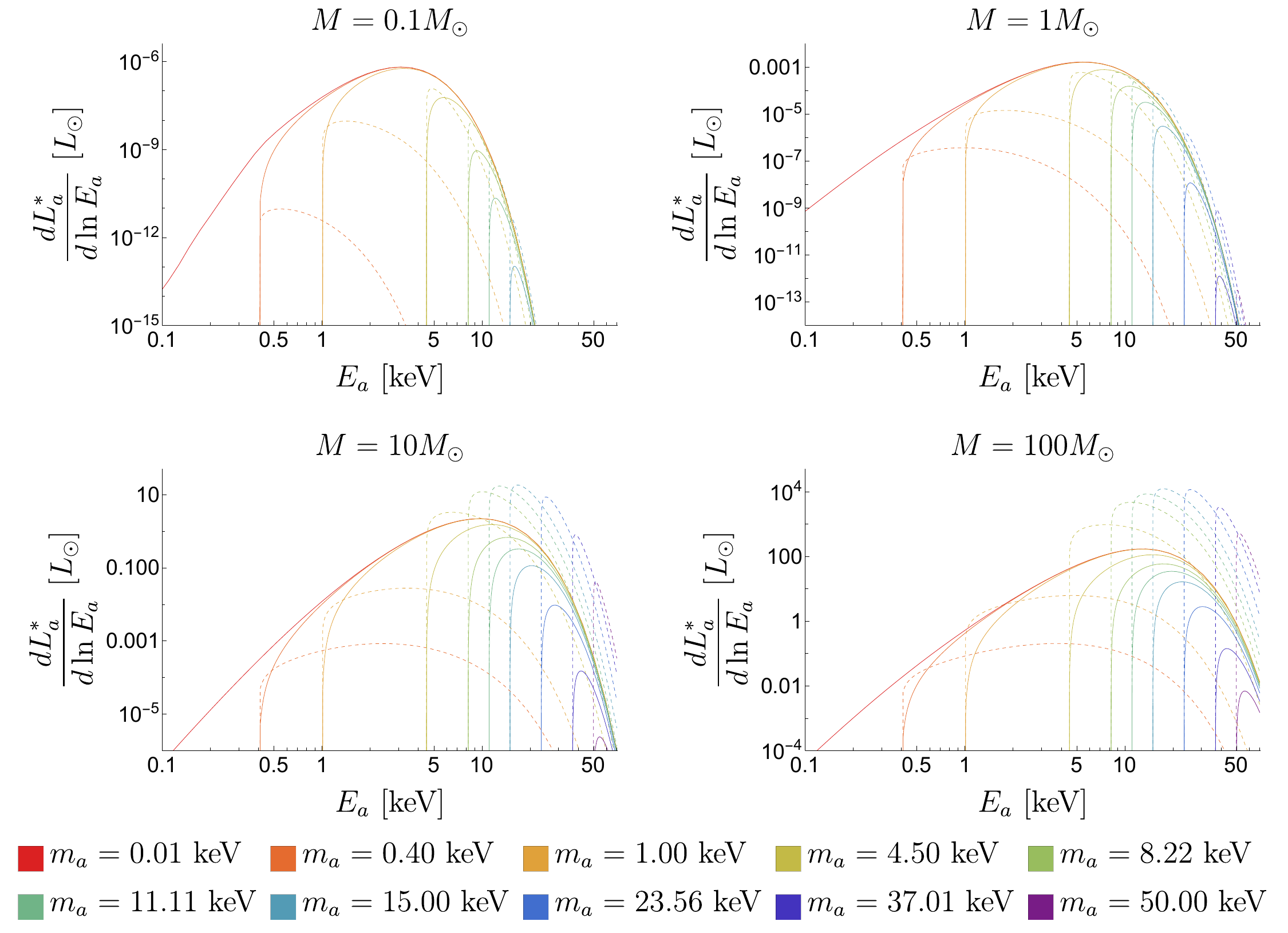}
  \caption{Axion luminosity per unit logarithmic energy range $dL_a^*/d\ln E_a$ from single main sequence stars of different masses $M$, for axions with $g_{a\gamma\gamma}=10^{-10}\GeV^{-1}$ and different masses $m_a$. Shown in solid and dot-dashed lines are the $dL_a^*/d\ln E_a$ from the Primakoff effect (Eq.\eqref{Primakoff}) and photon coalescence (Eq.\eqref{Coalescence}), respectively. The core temperatures of the assumed $0.1 M_\odot$, $M_\odot$, $10 M_\odot$, and $100 M_\odot$ model stars are about $0.8\keV$, $1.6\keV$, $3.0\keV$, and $4.4\keV$, respectively. }
  \label{fig:dLadEa}
\end{figure}

We compute the axion production using the full stellar profiles from MESA, but also provide simple fits to these stellar profiles in Appendix.~\ref{AppendixA}, which can be used for reproducing our results or for other purposes. In Fig.~\ref{fig:dLadEa}, we show the results of the integration over stellar layers \eqref{Nasingle} in terms of the energy distribution of the axion luminosity, $dL_a^*/d\ln E_a=E_a^2d\dot{N}_a^*/dE_a$, for several representative stellar masses. As one would expect, almost all the axions from a star are emitted from the core region of the star. The axion emission from our solar-mass star is close to that from the Sun \cite{CAST:2007jps} though with small differences, due to the slightly different chemical composition and stellar age assumed. Our results show that axion production from photon coalescence tends to dominate over that from the Primakoff process for massive stars and sufficiently high axion masses.

To better understand how the axion emission changes with stellar mass, we next derive rough scaling laws of the axion luminosity $L_a^*(M)$ with the stellar mass $M$ using our MESA fits (see Appendix.~\ref{AppendixA}). In the remainder of this sub section, we focus on stars with masses $M_\odot \leq M\leq 100M_\odot$ for which our MESA fits work well, consider axions with masses $m_a\gtrsim \keV$ for which the plasma masses in stellar cores\footnote{The core plasma mass varies over the stellar ensemble in the range $\omega_{\rm p,c}\sim 0.3-0.04\keV$.} are negligible ($m_a^2\gg \omega_{\rm p}^2$), and ignore log factors unless they have exponentially large arguments, which may occur in the expression for the axion production rate from photon coalescence \eqref{Coalescence} due to the sinh functions. As shown in Appendix.~\ref{AppendixA}, the temperature profiles in the cores of our MESA-generated stars are well fitted by the exponential profile $T(r)=T_c e^{-r/r_T}$. Using the characteristic length scale of the temperature profile $r_T$ as a proxy for the stellar core radius, the axion luminosity from a star can be crudely estimated as 
\begin{align}
    L_a^*\sim r_T^3\left(E_a \frac{d\dot{n}_a^{*}}{dE_a}\right)_{\rm peak}\Delta E_a  .  
\end{align}
The peak energy at which axion is sourced occurs at $E_{\rm peak}\sim \text{max}\left[3T_c,m_a\right]$ for the Primakoff process and $E_{\rm peak}\sim m_a$ for photon coalescence, while the width of the peak for both processes is set by the Boltzmann factor, i.e. $\Delta E_a\sim T_c$. Hence, the axion luminosity from the Primakoff effect and photon coalescence scale as
\begin{align}
    \left.L_a^*\right|_{\rm Prim.}&\propto g_{a\gamma\gamma}^2r_T(M)^3 \kappa_c(M)^2T_c(M)^2\left[\text{max}\left(3T_c(M),m_a\right)\right]^3e^{-m_a/T_c(M)},\label{eq:Lprimapprox}\\
    \left.L_a^*\right|_{\rm coal.}&\propto g_{a\gamma\gamma}^2r_T(M)^3m_a^5T_c(M)^2e^{-m_a/T_c(M)}.\label{eq:Lcoalapprox}
\end{align}
According to our MESA fits for the stellar mass range $1-100 M_\odot$, the core temperature $T_c(M)$, core radius $r_T(M)$, and inverse screening length $\kappa_c(M)$ scale as $T_c\propto M^{0.22}$, $r_T\propto M^{0.61}$, and $\kappa_c\propto M^{-0.76}$. These give
\begin{align}
    \left.L_a^*\right|_{\rm Prim.}&\propto\begin{cases} 
    g_{a\gamma\gamma}^2M^{0.75+b(M)}, &M\lesssim M_b\\
    g_{a\gamma\gamma}^2M^{1.41}, &M\gtrsim M_b
    \end{cases}, \label{primscalings}\\
    \left.L_a^*\right|_{\rm coal.}&\propto \begin{cases}
    g_{a\gamma\gamma}^2 M^{2.27+b(M)}, &M\lesssim M_b\\
    g_{a\gamma\gamma}^2M^{2.27}, &M\gtrsim M_b
    \end{cases},\label{coalscalings}
\end{align}
where
\begin{align}
    b(M)=\frac{1-(M/M_\odot)^{-0.22}}{\ln (M/M_\odot)} \frac{m_a}{1.83\keV},\quad\quad M_b=\left(\frac{m_a}{1.83\keV}\right)^{4.54}M_\odot . \label{bMb}
\end{align}
The exponent $b(M)$ captures the exponential suppression from the Boltzmann factor $e^{-m_a/T_c(M)}$ which is important for $M\lesssim M_b$ (for which $m_a\gtrsim T_c(M)$) and is a monotonically decreasing function of $M$ which varies in the range $(0.22-0.14)\times m_a/1.83\keV$ as the stellar mass is varied from $M_\odot$ to $100 M_\odot$.

\subsection{Stellar Axion Background}\label{s:StAB}

The aggregate of all the stars in the universe can source a cosmic population of axions which we refer to as the \textit{Stellar Axion Background} (StAB). Let us begin with a quick estimate for the largest possible energy density of the StAB. The limits on the axion-photon coupling $g_{a\gamma\gamma}\lesssim 6.6\times 10^{-11}\GeV$ for $m_a\lesssim 10 \keV$ from CAST and globular cluster observations allow a Sun-like (near solar mass, main sequence) star to emit axions with luminosity $\lesssim 10^{-3} L_\odot$ \cite{CAST:2007jps}. This can be linked to the cosmic optical background (COB), which is dominantly sourced by Sun-like stars \cite{Madau:2014,Gardner:1997js,1998ApJ...498..106M}. The observed energy density of the COB $\rho_{\rm COB}\sim 10^{-5}-10^{-4}\meV^4$ \cite{Hill:2018trh} sets a rough upper limit on the stellar axion background energy density from Sun-like stars
\begin{align}
    \rho_{\rm StAB}^{M\sim M_\odot}\lesssim 10^{-3}\rho_{\rm COB}\sim 10^{-8}-10^{-7}\meV^4    .
\end{align}
A similar estimate for the maximum $\rho_{\rm StAB}^{M\sim M_\odot}$ can be obtained from the luminosity density of the universe, which has been measured to be $2\times 10^8 L_\odot/\text{Mpc}^3$ around the optical band ($E_\gamma\sim 3\eV$) \cite{Fukugita:1997bi,2dGRS:2001lwv,SDSS:2002vxn}, implying that the cosmic density of Sun-like stars is $\sim 10^8/\text{Mpc}^3$. Combining this and that the axion luminosity of a Sun-like star is at most $\sim 10^{-3}$ of its total luminosity we find
\begin{align}
    \rho_{\rm StAB}^{M\sim M_\odot}\lesssim \frac{10^8\text{ Sun-like stars}}{\text{Mpc}^3}\times 10^{-3} L_\odot\times H_0^{-1}\sim  10^{-7}\text{ meV}^4 .
\end{align}
If a substantial fraction of the StAB decays to photons, it can leave an imprint in the cosmic X-ray background (CXB) spectrum, which has been measured to have an energy density of $\rho_{\rm CXB}\sim 10^{-8}\meV^4$ in the $1-10 \keV$ energy range. The above estimates suggest that there is a potential for probing the axion parameter space below the globular cluster bound with the CXB spectrum, which will depend on not only the spectral shape of the StAB decay signal but also how well the CXB spectrum is measured and understood.

\begin{figure}
  \centering
  \includegraphics[width=0.95\linewidth]{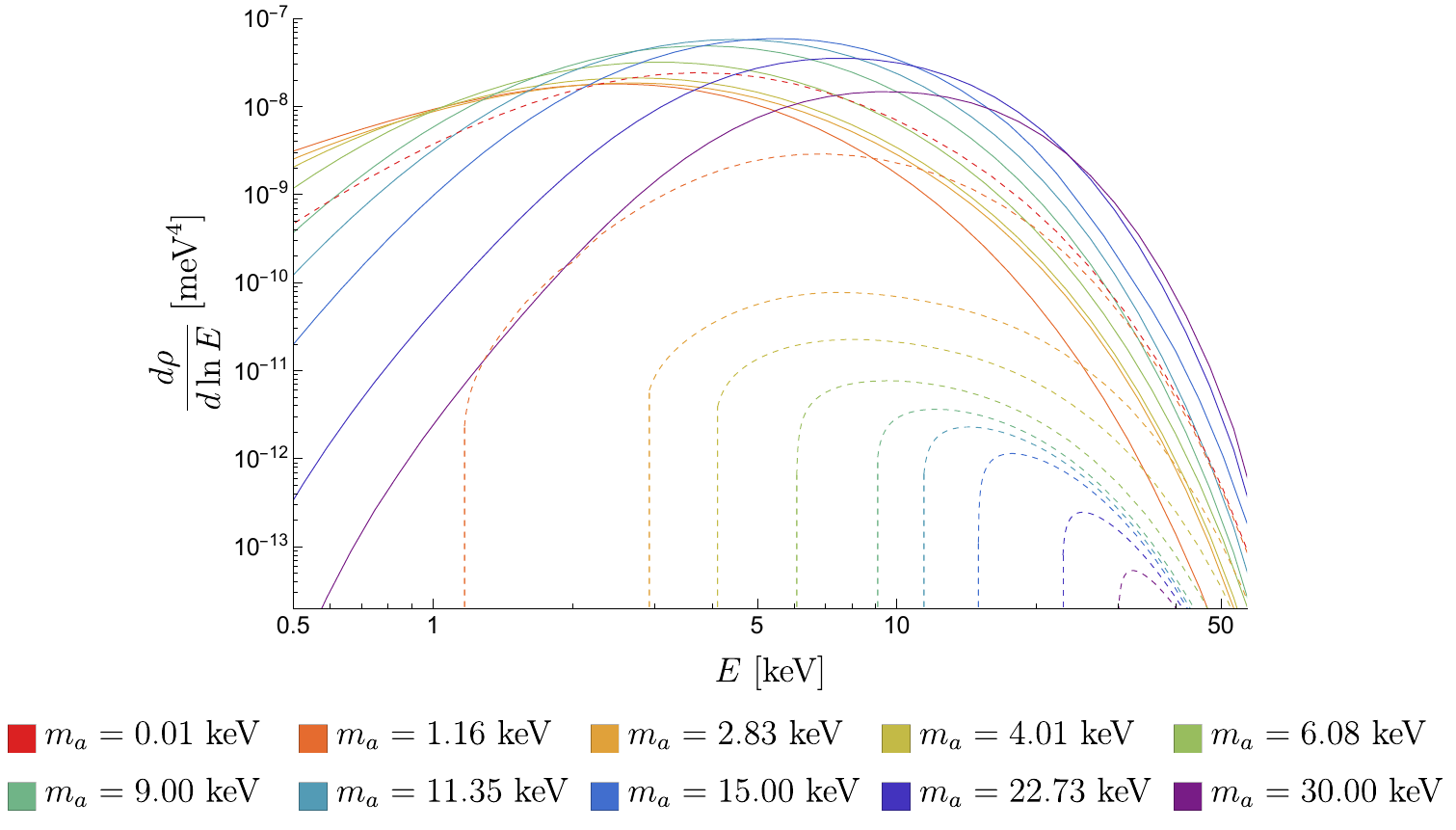}
  \caption{The spectra of energy density per unit logarithmic energy interval $d\rho/d\ln E$ of \textit{StAB (dashed)} and \textit{StAB-decay photons (solid)} at the current epoch ($z=0$) for $g_{a\gamma\gamma}=10^{-10}\GeV^{-1}$ and different axion masses $m_a$. As $m_a$ is increased from a small value, the StAB spectrum becomes progressively suppressed due to shorter axion decay lifetime, kinematic energy cut ($E>m_a$), and Boltzmann suppression. The photon spectrum increases at the start due to increased fraction of decayed axion, but decreases at higher $m_a$ once Boltzmann suppression of the axion production kicks in for most of the stars. The widths of the StAB and StAB-decay photon spectra are determined by many factors, with the general trend being that they are narrower for heavier axions due to the more severe kinematic cuts of the axion production and the smaller number of massive stars with sufficiently high core temperatures to efficiently source axions.}
  \label{fig:FinalSpectrum1}
\end{figure}

\begin{figure}
  \centering
  \includegraphics[width=0.95\linewidth]{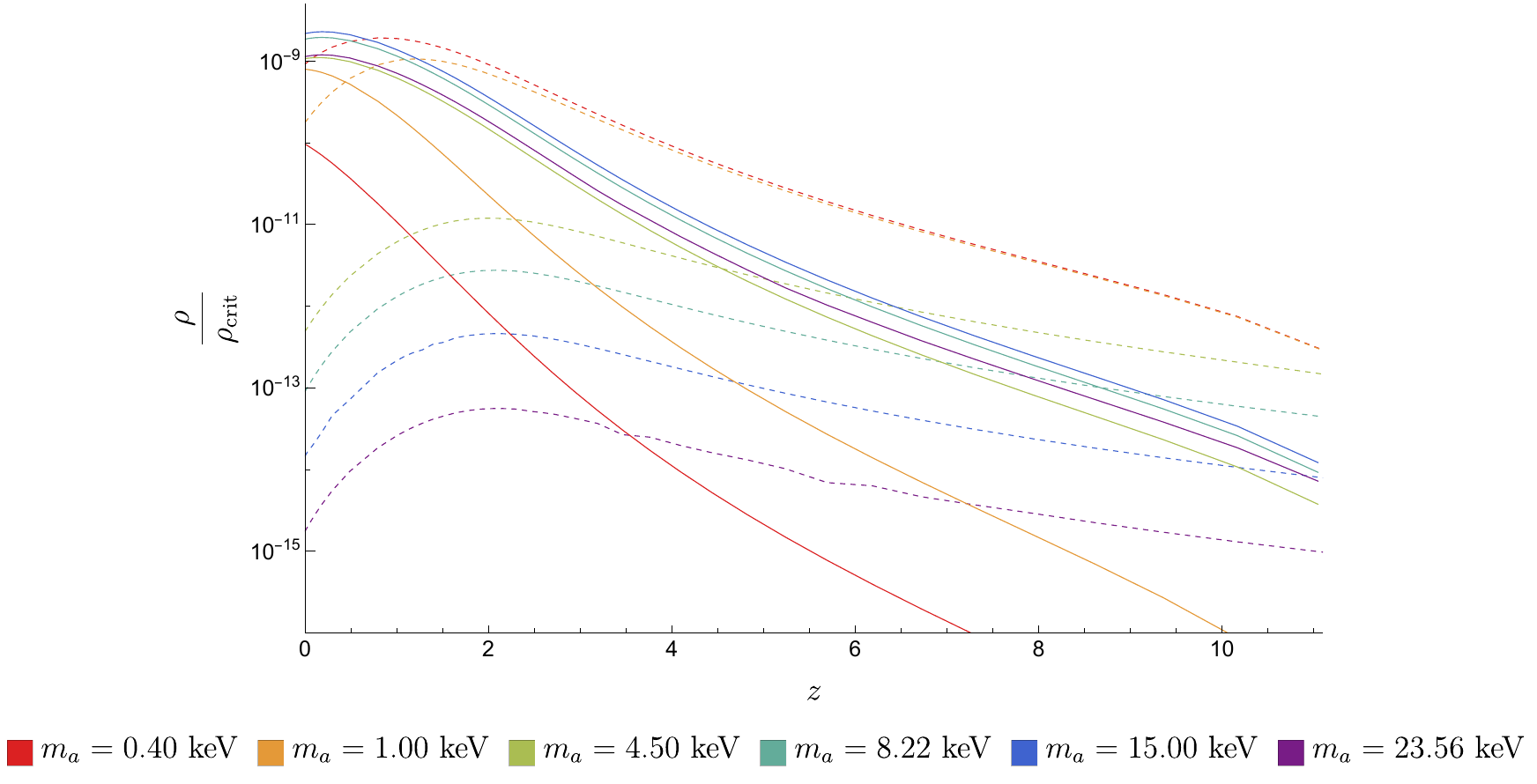}
  \caption{The redshift evolution of the physical energy 
  density of \textit{StAB (dashed)} and \textit{StAB-decay photons (solid)} relative to the critical density of the universe $\rho_{\rm crit}= 3M_P^2 H(z)^2$ for $g_{a\gamma\gamma}=10^{-10} \text{GeV}^{-1}$ and different axion masses $m_a$. For low axion masses, the StAB abundance accumulates over time while a small fraction of it gradually leaks into photons. For high axion masses, the sourced axions decay promptly into photons, and consequently the StAB energy density is set by the quasi-equilibrium between its sourcing and decay. The StAB energy density in this case is proportional to and reflects the redshift dependence of the comoving star formation rate $\psi(z)$ (see Fig.~\ref{fig:massfunctionandformationrate}), while the StAB-decay photons inherits essentially all the energy of the sourced axions and accumulates over time.
}
  \label{fig:FinalSpectrum2}
\end{figure}

In the remainder of this subsection, we compute the spectra and evolution of the StAB more carefully. The resulting StAB and StAB-decay photon spectral energy density at the present epoch ($z=0$) and the redshift evolution of their total energy densities are shown in Figs.~\ref{fig:FinalSpectrum1} and \ref{fig:FinalSpectrum2}, for different axion masses. We consider only axion emissions from main sequence stars, neglect the metallicity- and time-dependence of the axion production rate from a single star, and neglect the backreaction of axion emission on stellar evolution. While there are many sources of uncertainties associated with the properties and distribution of stars, we find that the dominant axion sourcing occurs at redshifts $z\lesssim 2$ where the star formation rate is well established.

We now proceed with the full calculation of the StAB spectrum and evolution. The StAB \textit{comoving} energy spectrum is given by
\begin{align}
    \frac{d\rho_a}{d\ln E_a}\left(E_a,z\right)=E_a^2 \frac{dn_a}{dE_a}\left(E_a,z\right),
\end{align}
with the comoving spectral axion density $dn_a/dE_a$ evolving as
\begin{align}
    -\frac{d^2n_a}{dE_a dz}(E_a,z)=\frac{1}{H(z)(1+z)}\left[\frac{d\dot{n}_a^{\text{all }\star}}{dE_a}(E_a, z)-\Gamma_{a\rightarrow \gamma\gamma}(E_a)\frac{dn_a}{dE_a}(E_a, z)\right],
\end{align}
where $H(z)$ is the Hubble rate at redshift $z$ and $\Gamma_{a\rightarrow \gamma\gamma}(E_a)$ is the decay rate of axion in the rest frame. The axion number spectrum per comoving volume observed at redshift $z$ is found by solving the above differential equation. We express this solution as an integral over the previous epochs 
\begin{align}
    \frac{dn_a}{dE_a}(E_a,z)&=\int_{z}^{\infty}\frac{dz'}{H(z')(1+z')}\frac{dE_a^\prime}{dE_a}\frac{d\dot{n}_a^{\text{all }\star}}{dE_a^\prime}(E_a^\prime, z')e^{-\int_{z}^{z'}\frac{dz''}{H(z'')(1+z'')}\Gamma_{a\rightarrow\gamma\gamma}(E_a^{\prime\prime})},
\end{align}
where the exponential factor is the fraction of axions that do \textit{not} decay to photons between redshift $z$ and $z'$ ($z'>z$), $E_a^\prime$ is the axion energy when it is emitted at redshift $z'$,
\begin{align}
    E_a^\prime=\sqrt{m_a^2+\left(\frac{1+z'}{1+z}\right)^2(E_a^2-m_a^2)},
\end{align}
$E''$ is defined similarly but with $z'\rightarrow z''$, and $d\dot{n}_a^{\text{all }\star}/dE_a^\prime(E_a',z')$ is the total axion production rate per unit emission energy $E_a^\prime$ per unit comoving volume produced by all stars present at redshift $z'$ which is given by
\begin{align}
    \frac{d\dot{n}_a^{\text{all }\star}}{dE_a}(E_a,z)&=\int_{M_{\rm min}}^{M_{\rm max}}\frac{dM}{M_\odot}\,\phi(M)\int_{0}^{t(z)}dt'\,\psi[z(t')]\frac{d\dot{N}_a^{\star}}{dE_a}(E_a, M, t(z)-t')\nonumber\\
    &=\int_{M_{\rm min}}^{M_{\rm max}}\frac{dM}{M_\odot}\,\phi(M)\frac{d\dot{N}_a^{\star}}{dE_a}(E_a,M)\int_{\text{max}[0,t(z)-t_{\rm life}(M)]}^{t(z)}dt'\,\psi[z(t')],
\end{align}
where $\phi(M)$ is the normalized stellar initial mass function for an assumed stellar mass range $[M_{\rm min},M_{\rm max}]$, $\psi(z)$ is the comoving star formation rate density at a given redshift $z$, $t_{\rm life}(M)$ is the main-sequence lifetime of a star of a given mass $M$\footnote{The assumed values of $t_{\rm life}(M)$ can be found in Appendix.~\ref{AppendixA}.} and $d\dot{N}_a^{\star}/dE_a$ is the axion production rate per unit emission energy $E_a$ produced by a single star, given by Eq.\eqref{Nasingle}. We have assumed in going to the second line that $d\dot{N}_a^{\star}/dE_a$ is constant and has support only during the main-sequence phase of a star
\begin{align}
    \frac{d\dot{N}_a^{\star}}{dE_a}(E_a,M,t-t')=\frac{d\dot{N}_a^{\star}}{dE_a}(E_a,M)\Theta\left[t_{\rm life}(M)-\left(t-t'\right)\right],
\end{align}
where $t-t'$ is the age of the star.

The cosmic star formation rate density $\psi(z)$ is the total mass of stars formed per unit time per unit comoving volume at redshift $z$, typically written in units of $M_\odot/\text{yr}/\text{Mpc}^3$. We use the simple parameterization by Madau and Dickinson (2014) \cite{Madau:2014} and updated by Madau and Fragos (2017)\cite{Madau:2017},
\begin{align}
    \psi(z)=0.01\frac{(1+z)^{2.6}}{1+[(1+z)/3.2]^{6.2}}\ M_\odot\text{yr}^{-1}\text{Mpc}^{-3}.\label{starformationrate}
\end{align}
Note that most of the stars are produced at $z\approx 2$, where the star formation rate $\psi(z)$ is peaked. The above redshift-dependence of the star formation rate must be combined with the expansion history of the universe. We assume the standard $\Lambda$CDM model cosmology with the Hubble rate $H(z)$ evolving with redshift $z$ as,
\begin{align}
    H(z)=H_0\sqrt{\Omega_{\rm m}(1+z)^3+\Omega_\Lambda},
\end{align}
and take $H_0=70\text{ km/s/Mpc}$, $\Omega_{\rm m}=0.3$, and $\Omega_\Lambda=0.7$.

The initial mass function $\phi(M)\propto dN_*/dM$ characterizes the relative abundances of stars of different masses $M$ at birth. It is conventionally normalized such that, 
\begin{align}
    \int_{M_{\rm min}}^{M_{\rm max}}dM\,M\phi(M)=M_\odot   . 
\end{align}
The commonly used Salpeter initial mass function $\phi(M)\propto M^{-2.35}$ works well only for $M\gtrsim 0.5 M_\odot$. A more recent fit to various luminosity density data by Baldry and Glazebrook (2003) gives \cite{Baldry},
\begin{align}
    \phi&=\phi_0\begin{cases}
        \displaystyle \left(\frac{M}{0.5 M_\odot}\right)^{-1.5}, &M\leq 0.5 M_\odot\\
        \displaystyle \left(\frac{M}{0.5 M_\odot}\right)^{-2.2}, &M\geq 0.5 M_\odot
    \end{cases},\label{initialmassfunction}
\end{align}
where the prefactor $\phi_0$ is determined by the above normalization condition. The minimum and maximum stellar mass that we consider are $M_{\rm min}=0.1 M_\odot$ and $M_{\rm max}=100 M_\odot$. Stars with $M<0.1M_\odot$ do not ignite hydrogen burn and so do not enter the main sequence phase, while stars with $M>100 M_\odot$ are extremely rare. 

\begin{figure}
    \centering
    \includegraphics[width=0.9\linewidth]{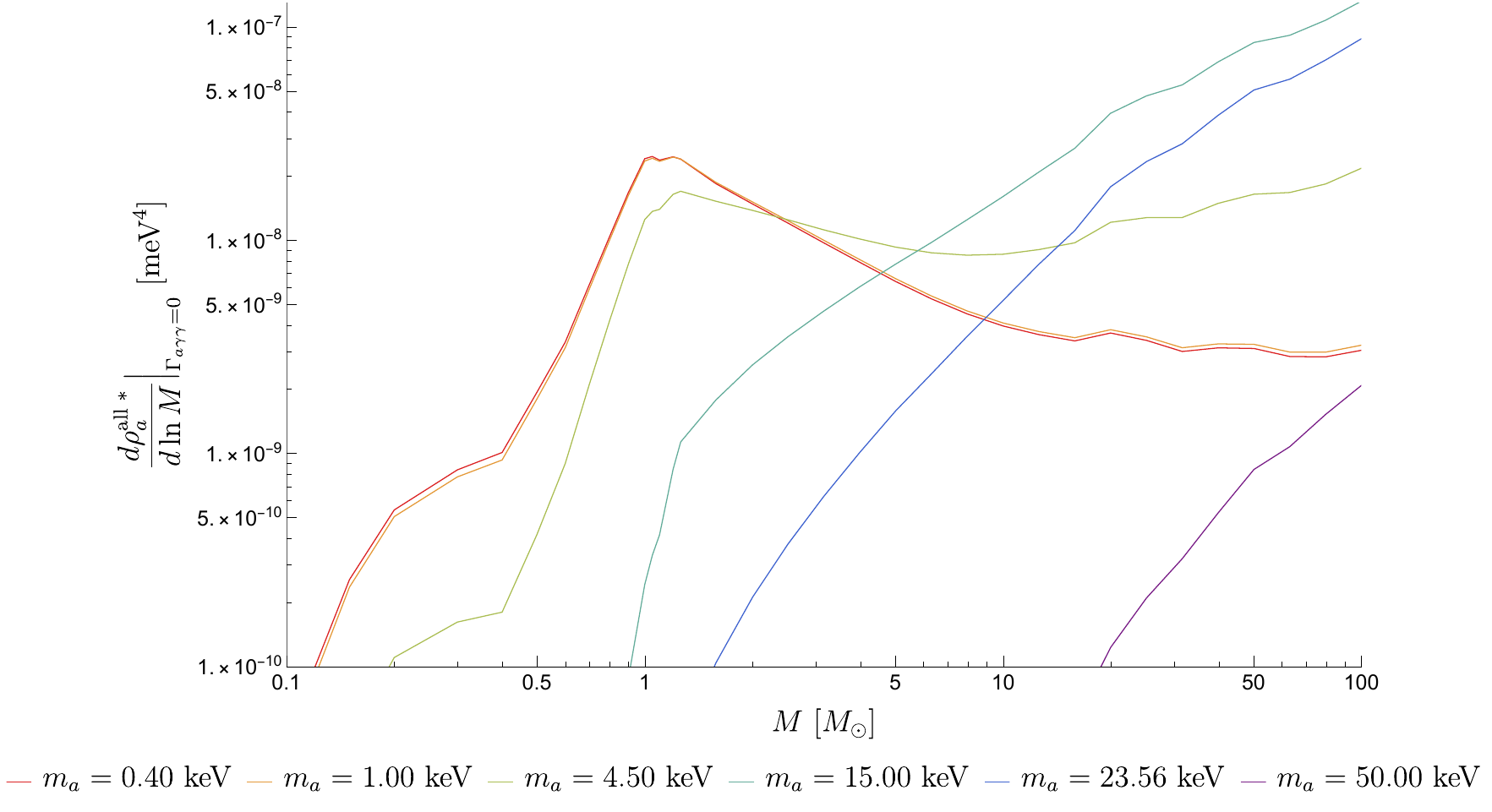}
    \caption{The StAB energy density per logarithmic stellar mass intervals in the absence of axion decays as a function of stellar mass $M$.}
    \label{fig:drhodlnM}
\end{figure}

To understand the relative importance of Main Sequence stars of different masses in sourcing axions, we compute the contribution to the present-day StAB energy density per logarithmic stellar mass intervals in the absence of axion decays as a function of stellar mass $M$, $\left.d\rho_a^{\text{all *}}/d\ln M\right|_{\Gamma_{a\gamma\gamma}=0}$. The results are displayed in Figure~\ref{fig:drhodlnM}, which shows that the cosmic axion sourcing can be dominated by either solar-mass stars ($M\sim M_\odot$) or the heaviest stars ($M\sim M_{\rm max}$) depending on the axion mass $m_a$ under consideration. For axion masses $m_a\lesssim 4\keV$, solar-mass stars dominate the axion sourcing, meaning that the StAB energy density and its decay products are not sensitive to the choice of $M_{\rm min}$ and $M_{\rm max}$. Whereas, for $m_a\gtrsim 4\keV$ the heaviest stars become the dominant source, and consequently our results in this heavy-axion case depend on the choice of $M_{\rm max}$. While there is thus far no generally accepted upper limit for the initial stellar mass function, upper mass limits of $100-200 M_\odot$ are typically assumed \cite{Baldry, Figer:2005gr,Oey:2005mn}. Our chosen value of $M_{\rm max}=100 M_\odot$, being on the lower end of this rough range, yields conservative estimates for the energy density of StAB and its decay photons, as well as the axion limits that we will derive in the next Section. We note that while the axion sourcing by post-main-sequence stars is beyond the scope of this paper, our finding that the heaviest main-sequence stars, despite their rarity, can dominate the sourcing of axions with $m_a\gtrsim 4\keV$ suggests the possibility of post-main-sequence stars being even stronger cosmic axion sources than main-sequence stars.

\begin{figure}
    \centering
    \includegraphics[width=\linewidth]{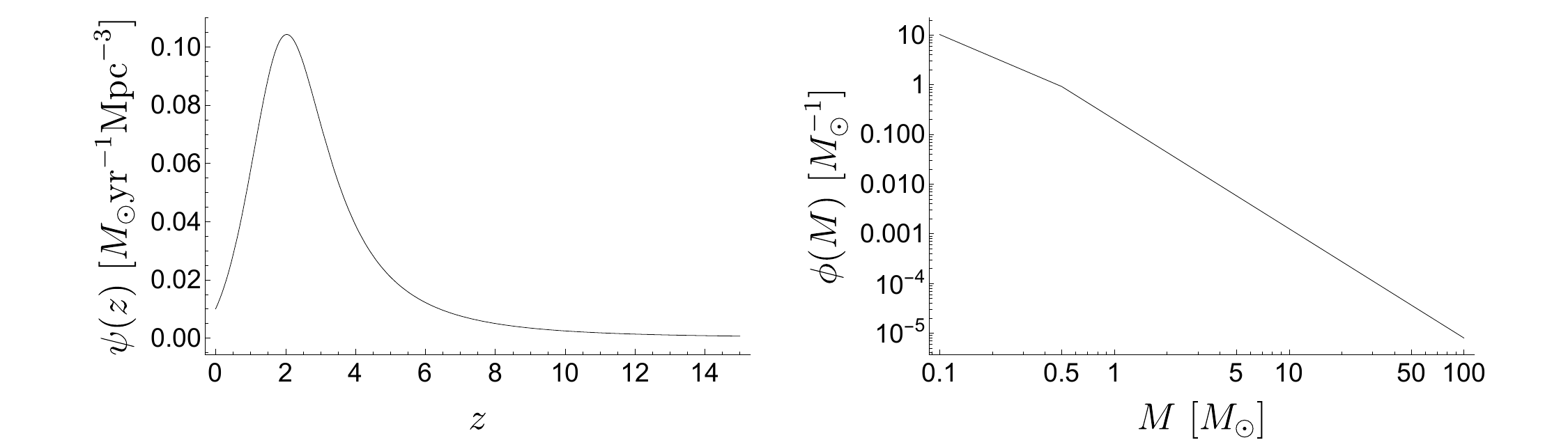}
    \caption{\textit{Left:} comoving star formation rate density as a function of redshift $\psi(z)$ (Eq.\eqref{starformationrate}) from \cite{Madau:2017}. \textit{Right:} stellar initial mass function $\phi(M)$ (Eq.\eqref{initialmassfunction}) from \cite{Baldry}.}
    \label{fig:massfunctionandformationrate}
\end{figure}

\section{X-ray from stellar axion decay}
\label{s:StABXray}

\subsection{Limits from the extragalactic X-ray background} 

Axions can spontaneously decay to two photons over the age of the universe with significant probabilities if their mass is relatively heavy, $m_a\sim \keV$.\footnote{For much lighter axions (lighter than or comparable to the plasma frequencies of the relevant media) efficient axion-photon conversions can also occur with the help of cosmic \cite{Mirizzi:2009nq,Mukherjee:2018oeb}, cluster \cite{Reynolds:2019uqt, Angus:2013sua, Conlon:2013txa}, galactic \cite{Xiao:2020pra,Dessert:2020lil}, or stellar \cite{Guarini:2020hps,Chang:2023quo,Dessert:2019sgw,Dessert:2022yqq,Chelouche:2008ta,OHare:2020wum,Caputo:2020quz,Fortin:2018ehg,Fortin:2021sst} magnetic fields. Given the considerable uncertainties in our knowledge of these magnetic fields we choose to focus on the spontaneous decay signals of the StAB.} The time-dilated axion decay lifetime in the cosmic frame is given by
\begin{align}
    \Gamma_{a\gamma\gamma}^{-1}=\frac{64\pi E_a}{g_{a\gamma\gamma}^2m_a^4}=0.14 t_{\rm U}\left(\frac{g_{a\gamma\gamma}}{10^{-10}\GeV^{-1}}\right)^{-2}\left(\frac{m_a}{1 \keV}\right)^{-4}\left(\frac{E_a}{4.5\keV}\right),
\end{align}
where $t_U=13.7\text{ Gyr}$ is the age of the universe. All the StAB-decay photons share the same energy of $\tilde{E}_\gamma=m_a/2$ in the axion rest frame but are Lorentz boosted in the cosmic frame by $v_a=\sqrt{1-(m_a/E_a)^2}$ (corresponding to a Lorentz factor $\gamma_a=E_a/m_a$) 
\begin{align}    E_\gamma=\gamma_a\left(\tilde{E}_\gamma+v_a\tilde{E}_\gamma \cos\theta\right)= \frac{E_a}{2}\left(1+v_a\cos\tilde{\theta}_\gamma\right).
\end{align}
Thus, $E_\gamma$ ranges from $E_{\gamma,\rm min}=E_a(1-v_a)/2$ to $E_{\gamma,\rm max}= E_a(1+v_a)/2$, corresponding to the axion-frame photon emission angles $\tilde{\theta}_\gamma=\pi$ (backward emission) and $\tilde{\theta}_\gamma=0$ (forward emission), respectively. The decay photons being isotropic in the axion frame (i.e. have flat distribution over solid angles) and $dE_\gamma\propto d\cos\tilde{\theta}_\gamma\propto d\tilde{\Omega}_\gamma$ imply that the cosmic-frame photon energy distribution from a single axion is flat in the range $[ E_{\gamma,\rm min},E_{\gamma,\rm max}]$. The energy spectrum of the StAB-decay photons at energy $E_\gamma$ is therefore related to that of the parent axions as\footnote{We do not include in our calculation X-ray absorption effects of the StAB decay signal in the intergalactic medium and the Milky Way. This results in only $\lesssim 10\%$ attenuation of the signal in the $1-10\keV$ energy range of interest and hence is negligible at the level of precision we are aiming for \cite{Wilms:2000ez,1983ApJ...270..119M,1974ApJ...187...57F}.}
\begin{align}
    \frac{d\dot{n}_\gamma}{dE_\gamma}(E_\gamma,z)&=\int_0^{\infty} d E_a  \frac{dn_a}{dE_a}(E_a,z)\Gamma_{a\gamma\gamma}(E_a)\frac{2\Theta(E_{\gamma,\rm max}-E_\gamma)\Theta\left(E_{\gamma}-E_{\gamma,\rm min}\right)}{E_{\gamma,\rm max}-E_{\gamma,\rm min}}\nonumber\\
    &= \int_{E_\gamma+\frac{m_a^2}{4E_\gamma}}^\infty dE_a \frac{dn_a}{dE_a}(E_a,z)\frac{2\Gamma_{a\gamma\gamma}(E_a)}{\sqrt{E_a^2-m_a^2}}.
\end{align}
The photon yield at the present epoch is then given by
\begin{align}
    \frac{d\rho_\gamma}{d\ln E_\gamma}(E_\gamma,z=0)=E_\gamma^2\int_{0}^{\infty}\frac{dz'}{H(z')}\frac{d\dot{n}_\gamma}{dE_\gamma^\prime}(E_\gamma^\prime,z').
\end{align}

\begin{figure}
  \centering
  \includegraphics[width=\linewidth]{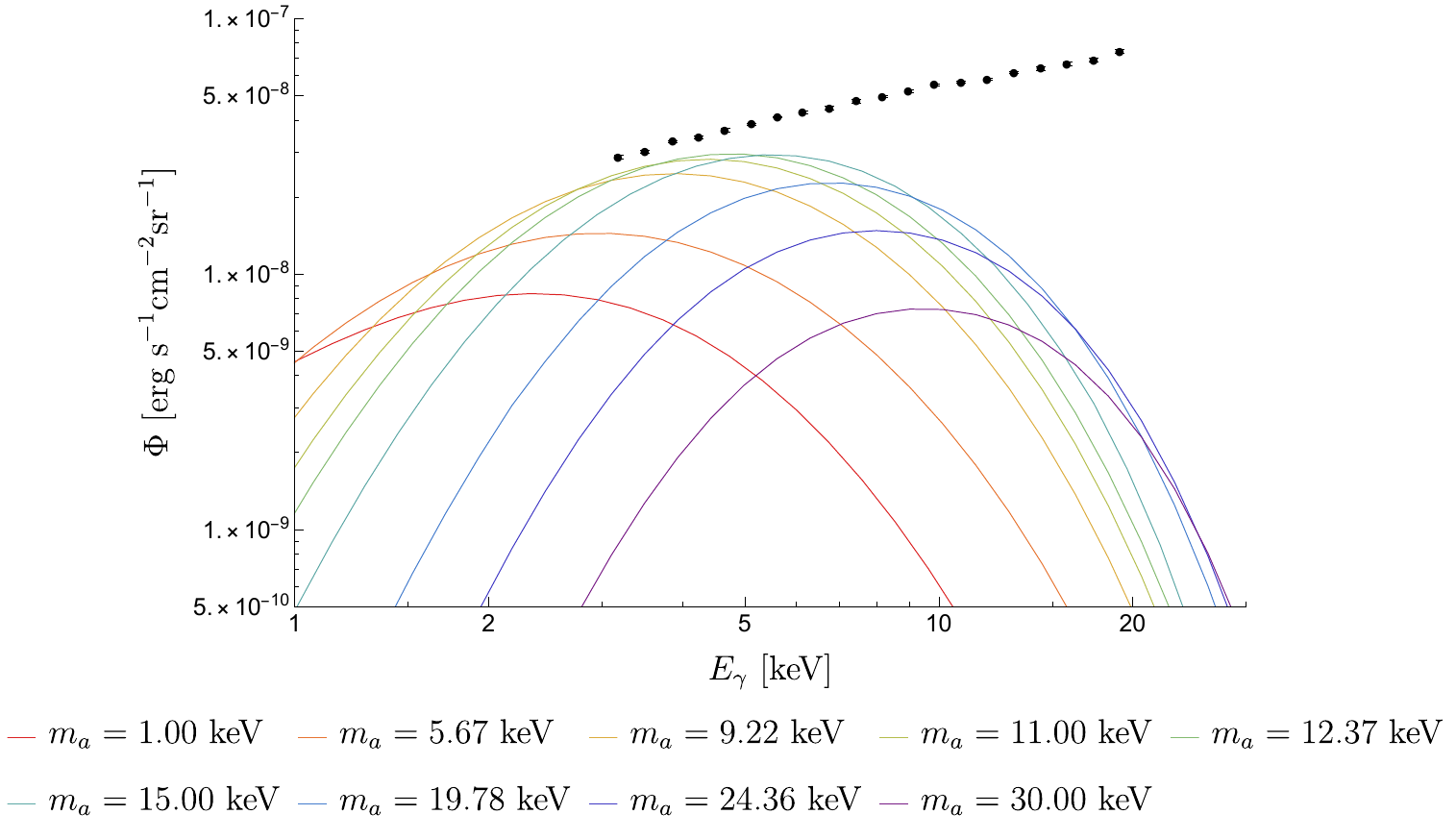}
  \caption{The StAB-decay photon spectral energy flux $\Phi$ (the local photon flux $F$ per unit logarithmic energy $\ln E$ interval per unit solid angle $\Omega$, $\Phi=d^2F/d\ln Ed\Omega$) for $g_{a\gamma\gamma}=10^{-10}\GeV^{-1}$ and different axion masses $m_a$. Also shown are the CXB spectrum data points from NuSTAR.
}
  \label{fig:XraySpectrum}
\end{figure}

The result is displayed in Figs.~\ref{fig:FinalSpectrum1},~\ref{fig:FinalSpectrum2}, and \ref{fig:XraySpectrum}, which show that the strongest StAB-decay signals occur in the parameter space where virtually all the star-produced axions decay to photon before the present epoch, in which case the StAB-decay photon energy density is completely determined by the total amount of axion energy sourced. The present day photon energy spectrum from StAB decay lies mainly in the X-ray regime and is most detectable in the $\sim 1-10 \keV$ energy range, corresponding to axions with masses $m_a\approx 0.5-30 \keV$. For the maximum axion-photon coupling compatible with the CAST and globular cluster bounds, $g_{a\gamma\gamma}= 6.6\times 10^{-11}\GeV^{-1}$, the total StAB-decay photon energy density can be as high as $\rho_\gamma\sim 10^{-8}\meV^4$ which amounts to X-ray fluxes per unit solid angle of $\sim 10^{-8}\text{ erg}\text{ s}^{-1}\text{cm}^{-2}\text{sr}^{-1}$, i.e. comparable to that of the observed CXB in the same energy range. Hence the StAB X-ray signal, if present, can be seen as a bump in the low-energy tail of the CXB spectrum which is known to peak at around 30 keV energy. As we will discuss in the next subsection, the known shape of the axion decay signal enables us to disentangle it from adequately-modeled backgrounds and thereby probe the existence of axion.

\begin{figure}
    \centering
    \includegraphics[width=0.95\linewidth]{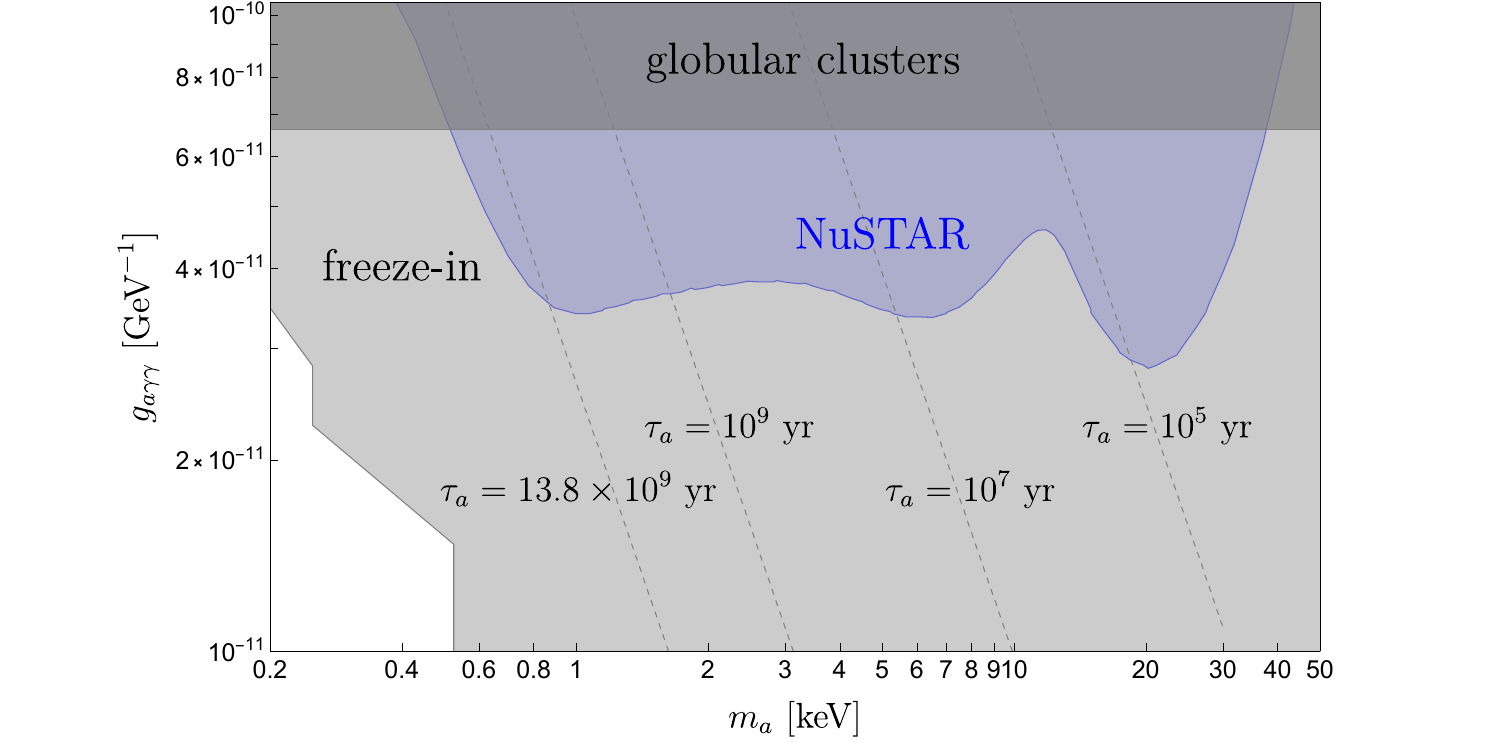}
    \caption{Exclusion limits on the axion parameter space at 95\% confidence level derived from the chi-squared goodness of fit test against the CXB data from NuSTAR. The axion decay lifetime $\tau_a$ is calculated at the peak energy of the StAB energy spectrum which is expected to be $\sim 4.5\ \rm keV$ for $m_a\leq 4.5\ \rm keV$ and at $\sim m_a$ for $m_a> 4.5\ \rm keV$. The light gray region is the constraint derived from the decay of relic axions produced via freeze-in from \cite{Langhoff:2022bij}.}
    \label{fig:limit1}
\end{figure}

Several generations of X-ray instruments such as Chandra, HEAO, NuSTAR, Swift-XRT, and XMM-Newton have measured the CXB in the $\sim 1-10 \keV$ energy range where the StAB decay signal is most likely to be found \cite{DeLuca:2003eu, Bauer:2004tk,Brandt:2005ra, 2012A&A...548A..87M,Harrison, 2012ApJ...752...46L,Hill:2018trh,Cappelluti:2017miu}. We adopt for our analysis the CXB data from NuSTAR \cite{Krivonos:2020qvl}. The observed CXB spectra should be interpreted as the sum of the axion decay signal and the astrophysical background, which is known to be primarily due to active galactic nuclei. We can probe an axion parameter space based on how the inclusion of the StAB X-ray predicted by that parameter space affects the quality of fit to the CXB data. We define the CXB spectral energy flux $\Phi_{\rm CXB}(E)=d^2F_{\rm CXB}/d\ln E d\Omega$ (with the units of $\text{erg}\text{ s}^{-1}\text{cm}^{-2}\text{sr}^{-1}$) as the CXB photon energy flux $F_{\rm CXB}$ per unit logarithmic energy interval per unit solid angle at energy $E$ and quantify the goodness of fit to the CXB data $\Phi_{\text{CXB}, i}$ with the following chi-squared function
\begin{align}
    \chi^2=\sum_{i} \frac{1}{\sigma_{\Phi_{\text{CXB}, i}}^2}\left(\Phi_{\text{CXB}, i}-\left.\Phi_{\rm CXB}^{\rm model}(g_{a\gamma\gamma},m_a,\boldsymbol{\theta}_{\rm bg})\right|_{E_i}\right)^2,
\end{align}
where the sum is over the energy $E_i$ of the X-ray telescope; $\Phi_{\text{CXB}, i}$ and $\sigma_{\Phi_{\text{CXB}, i}}$ are respectively the CXB spectral energy flux and its associated error at energy $E_i$. We model the spectral energy flux $\Phi(E)$ of the CXB as the sum of the expected signal from StAB decay and an attenuated power law model for the CXB background \cite{Krivonos:2020qvl,Gruber}
\begin{align}
    \Phi_{\rm CXB}^{\rm model}(g_{a\gamma\gamma},m_a,\boldsymbol{\theta}_{\rm bg})&=\frac{1}{4\pi}\left(\frac{d\rho_{\gamma}}{d\ln E}\right)_{\rm StAB}+A\left(\frac{E}{\keV}\right)^{2-\Gamma}e^{-E/E_0}, \label{CXBmodel}
\end{align}
where $\boldsymbol{\theta}_{\rm bg}=\{A,\Gamma,E_0\}$. For each axion mass $m_a$, we first minimize the $\chi^2$ over all the parameters other than the mass $m_a$ to obtain the best-fit chi-squared $\left[\chi^2(m_a)\right]_{\text{best}(g_{a\gamma\gamma}, \boldsymbol{\theta}_{\rm bg})}$. Then we calculate again the $\chi^2$ but now minimizing over only the background parameters $\boldsymbol{\theta}_{\rm bg}$, giving $\left[\chi^2(g_{a\gamma\gamma},m_a)\right]_{\text{best}(\boldsymbol{\theta}_{\rm bg})}$. By Wilk's theorem \cite{Cowan:2010js}, the difference of these two chi-squared values follow a chi-squared distribution for one degree of freedom. This allows us to infer the likelihood of obtaining a given value of $g_{a\gamma\gamma}$ and place the $95\%$ confidence-level exclusion limits on the axion parameter space based on the following criterion
\begin{align}
    \left[\chi^2(g_{a\gamma\gamma},m_a)\right]_{\text{best}(\boldsymbol{\theta}_{\rm bg})}-\left[\chi^2(m_a)\right]_{\text{best}(g_{a\gamma\gamma}, \boldsymbol{\theta}_{\rm bg},)}>\chi^2_{95\%},
\end{align}
where $\chi^2_{95\%}=2.71$.

In the absence of the axion ($g_{a\gamma\gamma}=0$), the NuSTAR data are fit reasonably well with the attenuated power law model \eqref{CXBmodel}.\footnote{Letting all three CXB model parameters $\{A,\Gamma,E_0\}$ run free yields a best-fit chi-squared per degree of freedom $\chi^2/\text{dof}=1.56/17$ for the full NuSTAR data set. Fixing $\Gamma=1.29$ and varying over $\{A,E_0\}$ gives $\chi^2/\text{dof}=1.73/18$ (full data set) and $\chi^2/\text{dof}=0.98/17$ (with the last data point at $E=20\keV$ excluded). These reproduce almost exactly the $\chi^2/\text{dof}$ values reported in \cite{Krivonos:2020qvl}.} For the purposes of deriving the axion limits, we fix $E_0=41.13\keV$ (corresponding to the best fit $E_0$ value from \cite{Gruber}), remove the outlier data point $E\simeq 20\keV$, and vary over the remaining model parameters $\{A,\Gamma\}$, yielding an acceptable best-fit chi-squared per degree of freedom of $\chi^2/\text{dof}=0.97/17$ in the absence of axion. When the axion signal is included, the data do not display significant preference toward axion of any $m_a$ and $g_{a\gamma\gamma}$. As shown in Fig.~\ref{fig:limit1}, our analysis rules out at 95\% confidence level a swatch of axion parameter space $(m_a,g_{a\gamma\gamma})$ slightly below the cooling bound. The limits that we found lie in a parameter space that is already ruled out, mainly by limits based on relic axions from the early universe \cite{Cadamuro:2011fd, Langhoff:2022bij} and partially by limits from gravitationally bound axions around the Sun \cite{DeRocco:2022jyq,Beaufort:2023zuj}. Nevertheless, our limits are based on different assumptions from that of these earlier works. The considerations presented here can in principle provide independent and complementary tests in less minimal extended-sector models \cite{DeRocco:2020xdt,Jaeckel:2006xm,Chakraborty:2020vec,Boddy:2014yra,March-Russell:2020nun, Chang:2022gcs,Arvanitaki:2021qlj} with possibly non-standard cosmological scenarios \cite{Allahverdi:2020bys}, e.g. where the effective field theory parameters are time varying \cite{Gan:2023wnp,Baldes:2019tkl,Banerjee:2019asa}.

\subsection{X-rays from gravitationally bound objects}

While the integrated photon signal from StAB decay over the cosmic history is approximately isotropic, the decay signal from newly produced axions trace to some degree the spatial distribution of stars at the present epoch. We expect such decay signals from smaller redshifts to be enhanced in the directions of high density of stars such as groups and clusters of galaxies. The X-ray in those directions are typically also stronger, which means each source needs to be studied on case by cases basis. Given the diversity of astrophysical objects in the universe one might be able to find objects for which there is a relative enhancement of the stellar axion decay signal over the background. We assess the prospect for setting more stringent limits on the axion parameter space below $g_{a\gamma\gamma}\approx 10^{-11}\GeV^{-1}$ with the X-ray observations of gravitationally bound objects. Our aim here is simply to identify potential directions for future in depth studies.

If most of the optical photons and the StAB share the same source, namely Sun-like stars,\footnote{As per Figure.~\ref{fig:drhodlnM}, we expect Sun-like stars to dominate the axion luminosity of a gravitationally bound object of interest for $m_a\lesssim 4\keV$ if the object's stellar mass function resembles that of the present-day cosmic stellar mass function.}  the StAB X-ray sky should be highly correlated with the visible sky if the axions decay immediately outside the stars that source them. The finite decay lifetime of these axions, however, allow them to traverse a typical distance of $\ell_a\sim v_a\gamma_a/\Gamma_{a\gamma\gamma}$ before decaying into a pair of photons, leading to a spatial smearing of the X-ray from StAB at scales smaller than $\ell_a$. In the parameter space slightly below the cooling bound with $g_{a\gamma\gamma}\approx 10^{-11}\GeV^{-1}$ and $m_a\lesssim 10\keV$, we have $\ell_a\gtrsim 1\text{ Mpc}$ which is always much longer than the size of a galaxy ($\sim 1-100\text{ kpc}$) and can be comparable to the size of a galaxy cluster ($\sim 1-10\text{ Mpc}$). A useful picture to have before we proceed is that the decay of the axions sourced by a star would take place dominantly in a spherical shell of radius $\sim \ell_a$ and thickness $\sim \ell_a$ around the star. For the smallest possible $\ell_a$, these decay shells can be contained in a cluster and in that case the decay photons would appear to originate from the cluster. The axion decay signals become increasingly smeared out as $\ell_a$ is increased and eventually their sum become almost indistinguishable from complete homogeneity and isotropy.

Depending on the axion mass, star-emitted axions can behave like warm dark matter or dark radiation. We find that in all cases that yield substantial axion decay flux, the typical axion Lorentz factor is $\gamma_a\sim 1$. Hence, for simplicity, we will assume in what follows that the axion decays are isotropic. Strong relativistic beaming ($\gamma_a\gg 1$) of the decay photon may occur for axions that are orders of magnitude lighter than the typical temperature of stellar cores $T_c\sim 1.5\keV$, however such scenarios are of less interest in terms of their X-ray signals because the axions would have lifetimes much longer than the age of the universe. Slower axions with non-relativistic velocities $v_a\lesssim 10^{-3}$ are produced in stars at phase-space-suppressed rates but may accumulate in gravitationally bound objects over long timescales as in \cite{VanTilburg:2020jvl,DeRocco:2022jyq,Beaufort:2023zuj,Bastero-Gil:2021oky}. Such gravitationally-trapped axions would produce line-like decay photon signals, and potentially lead to stronger limits on the axion parameter space depending how many axions can be trapped at a given time. The latter will depend on the stability timescale of axion orbits in a many-body gravitational potential, which is nontrivial and requires a dedicated study.

The present-day stellar mass function is strongly dominated by Sun-like stars whose axion luminosity is $L_a\sim 10^{-3}\left(g_{a\gamma\gamma}/6\times 10^{-11}\GeV^{-1}\right)^2L_{\odot}$.
Axions are produced at roughly this luminosity as long as they are sufficiently light to avoid Boltzmann suppression, i.e. $m_a\lesssim 10\keV$. We would like to estimate the X-ray flux from the decaying axion cloud around an object of stellar concentration, which could be a galaxy, a galaxy, a galaxy group, or a cluster. Assuming the optical luminosity $L_{\rm O}$ from that object is dominated by Sun-like stars, the axion energy density $\rho_a$ at a radial position $r$ away from the center of such an object can be estimated as
\begin{align}
    \rho_a\sim 10^{-3}\left(\frac{g_{a\gamma\gamma}}{6\times 10^{-11}\GeV^{-1}}\right)^2L_{\rm O}\frac{e^{-\ell(r)/\ell_a}}{\ell(r)^2},
\end{align}
where $\ell(r)\sim \text{max}\left(r,R\right)$ is the typical distance from the point of interest to an arbitrary point in the object and we have assumed $\ell_a\gtrsim 1\text{ Mpc}\gtrsim R$. The flux per unit solid angle from axion decay in that object is then given by the integral of $\rho_a$ along the line of sight distance $s$ weighted by the axion decay probability per unit length, $F_{a\rightarrow\gamma\gamma}\sim (1/4\pi) \int ds \rho_a(1-e^{-\ell/\ell_a})/\ell_a$, yielding
\begin{align}
    F_{a\rightarrow\gamma\gamma}\sim 10^{-3}\left(\frac{g_{a\gamma\gamma}}{6\times 10^{-11}\GeV^{-1}}\right)^2L_{\rm O}\int \frac{ds}{\ell_a} \frac{e^{-\ell/\ell_a}\left(1-e^{-\ell/\ell_a}\right)}{4\pi \ell^2} .
\end{align}
For objects in which we reside, the relevant $r$ is whichever radius that dominates the X-ray flux. For distant objects, the relevant $r$ will be set by the direction and the FoV of the X-ray telescope  we are using. Below we provide crude estimates for the maximum axion-induced X-ray flux per unit solid angle from various types of astrophysical objects
\begin{itemize}
    \item \textit{The Sun} 
    \begin{align}
        F_{a\rightarrow\gamma\gamma}^{\rm Sun}\sim \left(\frac{g_{a\gamma\gamma}}{6\times 10^{-11}\GeV^{-1}}\right)^2\frac{10^{-3}L_\odot}{4\pi \text{AU}^2} \frac{\rm AU}{\ell_a}\lesssim 3\times 10^{-9}\text{ erg}\text{ s}^{-1}\text{cm}^{-2}\text{sr}^{-1} .
    \end{align}
    Here, the flux is the strongest when the distance to the sun $r$ is minimized, i.e. at $r\sim \rm AU$, because the $1/r^2$ decrease in the axion density is stronger than the linear increase $\propto r$ in the decay probability.
    \item \textit{The Milky Way galaxy}
    \begin{align}
        F_{a\rightarrow\gamma\gamma}^{\rm MW}&\sim 10^{10}\text{ Sun-like stars}\times \left(\frac{g_{a\gamma\gamma}}{6\times 10^{-11}\GeV^{-1}}\right)^2 \frac{ 10^{-3}L_\odot}{4\pi(10\text{ kpc})^2} \frac{10\text{ kpc}}{\ell_a}\nonumber\\
        &\lesssim 1\times 10^{-8}\text{ erg}\text{ s}^{-1}\text{cm}^{-2}\text{sr}^{-1} .  
    \end{align}    
    \item \textit{Distant clusters}
        \begin{align}
        F_{a\rightarrow\gamma\gamma}^{\rm cluster}&\sim 10^3\text{ galaxies}\times 10^{10}\text{ Sun-like stars} \times \left(\frac{g_{a\gamma\gamma}}{6\times 10^{-11}\GeV^{-1}}\right)^2 \frac{ 10^{-3}L_\odot}{4\pi(1\text{ Mpc})^2}\frac{1\text{ Mpc}}{\ell_a}\nonumber\\
        &\lesssim 1\times 10^{-7}\text{ erg}\text{ s}^{-1}\text{cm}^{-2}\text{sr}^{-1}  .
    \end{align} 
    We assume that the solid angle of the cluster $\Omega_{\rm cluster}\sim (\text{Mpc}/d)^2$ is greater than and so cover the entire FoV of the instrument, meaning that the axion decay signal is not diluted. For Chandra and XMM-Newton, $\Omega_{\rm FoV}\sim 10^{-5}\text{ sr}$ \cite{Abazajian:2001vt,Boyarsky:2006zi}.
\end{itemize}
The above maximum fluxes were obtained by setting $g_{a\gamma\gamma}=10^{-11}\GeV^{-1}$ and the shortest axion decay length corresponding to the highest $m_a$ without significant Boltzmann suppression, $\ell_a\sim \text{Mpc}$. By comparison, the previously-obtained StAB flux for the $g_{a\gamma\gamma}$ that saturates the globular cluster limit is at the level of $\sim  10^{-8}\text{ erg}\text{ s}^{-1}\text{cm}^{-2}\text{sr}^{-1}$ (comparable to the observed isotropic CXB). Hence, the X-ray signals from the directions  of stellar concentration can be enhanced by not more than an order of magnitude relative to the StAB X-ray signal in the same solid angle. This is essentially because the overall X-ray signals from these directions are not determined by the highly enhanced axion density in gravitationally bound objects. They are instead determined by the (more diluted) column densities of axion in these directions, i.e. the axion density integrated over the line of sight distance. Since the X-ray background relevant to these regions are also enhanced (or at best comparable in the periphery of these objects \cite{Eckert:2016bfe,Ghirardini:2017apw,Boyarsky:2006zi}) relative to that of the isotropic CXB, we expect only marginal improvements on the axion limits from what we have found previously with the isotropic CXB.

\section{Conclusion}\label{s:conclusion}
We have computed the spectra of axions with coupling only to electromagnetism produced in the cores of main sequence stars with masses in the range $0.1-100M_\odot$ using the stellar profiles obtained from the stellar evolution code MESA. We then use these axion spectra to estimate the abundance, spectrum, and time-evolution of the diffuse axion background sourced by all the stars in the universe across cosmic histories. This axion background can subsequently decay into X-rays and contribute to the cosmic X-ray background. The decay-photon spectrum has a calculable characteristic spectral shape with a peak expected at either half the average thermal energy, $3T_c/2\sim 2\keV$, or half the axion mass $m_a/2$, corresponding to relativistic and non-relativistic decays, respectively.

We provide in Appendix \ref{AppendixA} simple exponential fits of the temperature, inverse screening length, and plasma mass as a function of radius which approximate well the core profiles of $1-100 M_\odot$ main sequence stars used in our analysis. These fits in conjunction with Eqs.\eqref{Nasingle}, \eqref{Primakoff}, and \eqref{Coalescence} allow one to estimate the axion spectrum produced in the core of any main sequence star whose mass lies in the aforementioned range. Our ensemble of benchmark stars can be made more realistic by considering effects of time-evolution, varied chemical compositions, rotations, and magnetic fields. It would be interesting to include post-main-sequence, and perhaps also population III stars, in the stellar ensemble as the core temperatures in some of these non-main-sequence stars can be considerably higher than those of the main-sequence stars. These stars can dominate the production rate of heavy axions due to the relative lack of Boltzmann suppression. The formalism we use for calculating the properties of the StAB and its decay signal can serve as a template for estimating the stellar background of other light dark sector particles such as dark photons and millicharged particles.\footnote{For particles that are produced in stars dominantly near their surfaces rather than in their cores \cite{Redondo:2013lna,Redondo:2015iea}, one would need to capture the near-surface properties of the stars more carefully. }

\section*{Acknowledgments}
We thank Kevin Langhoff for collaboration in the early stages of the project and Gautham Adamane Pallathadka, Peter Graham, David E. Kaplan, Xuheng Luo, Nadav Outmezguine, Surjeet Rajendran for useful discussions at various stages of the project. This work was supported by NSF Grant No. 2112699 and the Simons Foundation.

\clearpage
\newpage
\appendix

\section{Simple fits to stellar properties from MESA}\label{AppendixA}

\begin{figure}
    \centering
    \includegraphics[width=\linewidth]{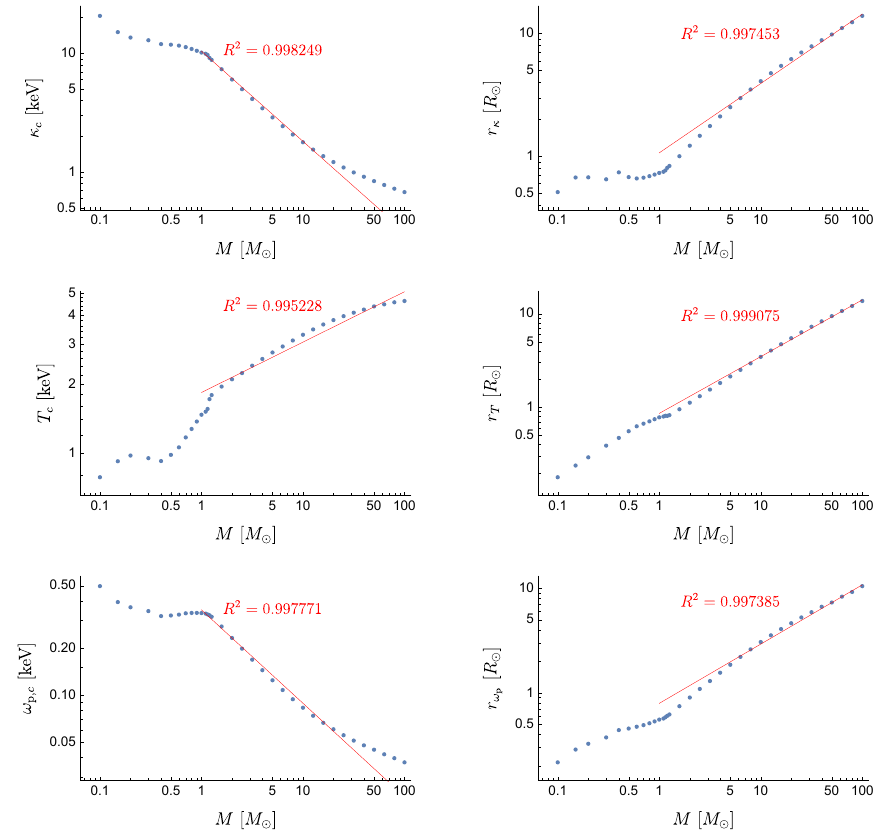}
    \caption{Visualisations of the fits in Eqs.\eqref{Tc}, \eqref{rT}, \eqref{kappac}, \eqref{rkappa}, \eqref{wc}, and \eqref{rw}. Blue dots are the data from MESA while the red lines are their power law fits.}
    \label{fig:parameters_fit}
\end{figure}

\begin{figure}
    \centering
    \includegraphics[width=\linewidth]{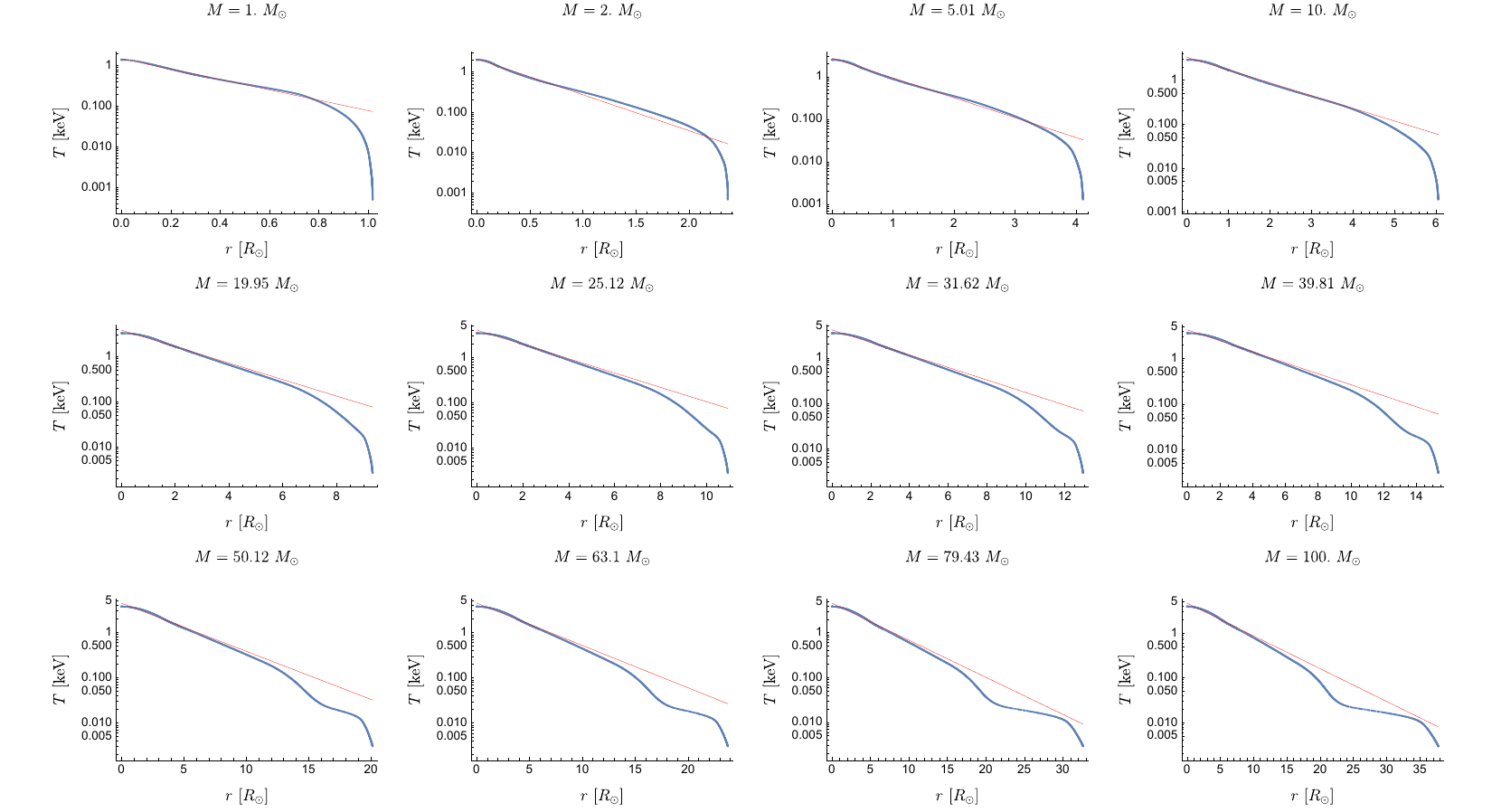}
    \caption{The temperature profiles of our MESA stars with masses $1-100 M_\odot$. Blue dots are the data from MESA while the red lines are their exponential fits as in Eq.\eqref{T}.}
   \label{fig:T_fit}    
\end{figure}

\begin{figure}
    \centering
    \includegraphics[width=\linewidth]{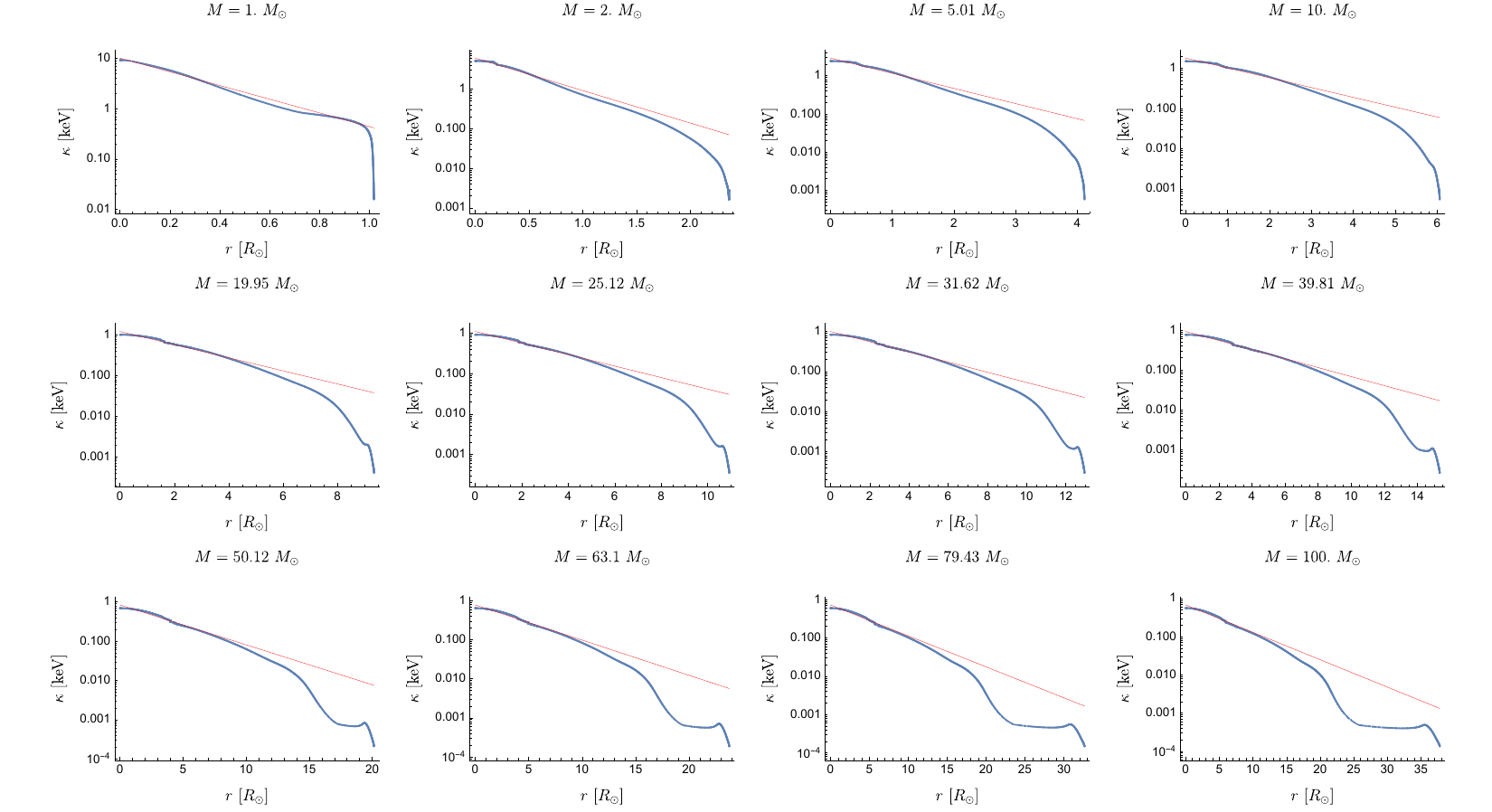}
    \caption{The inverse screening length profiles of our MESA stars with masses $1-100 M_\odot$. Blue dots are data from MESA while the red lines are their exponential fits as in Eq.\eqref{kappa}.}
   \label{fig:kappa_fit}
\end{figure}

\begin{figure}
    \centering
    \includegraphics[width=\linewidth]{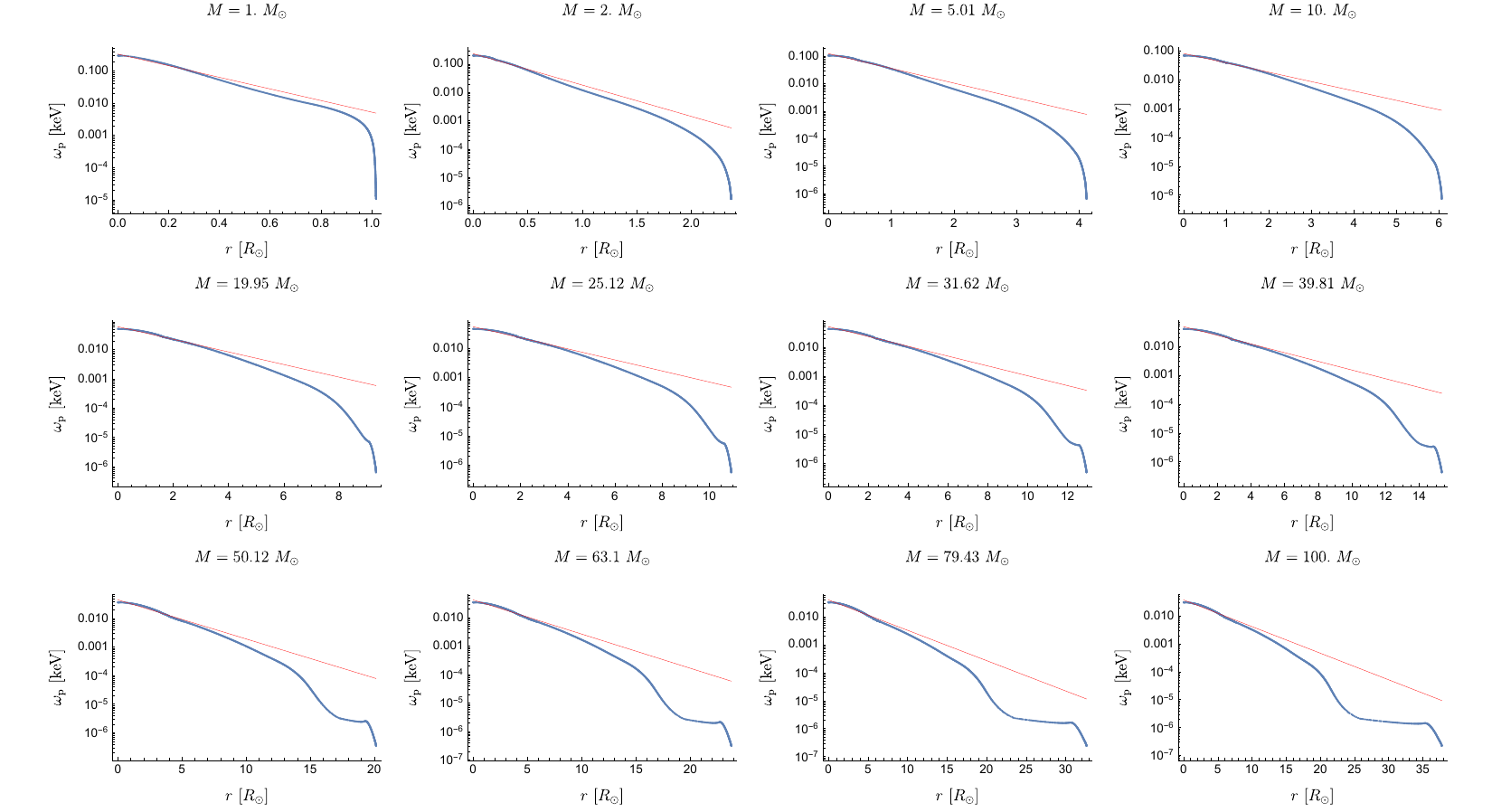}
    \caption{The plasma mass profiles of our MESA stars with masses $1-100 M_\odot$. Blue dots are the data from MESA while the red lines are their exponential fits as in Eq.\eqref{omega}. }
   \label{fig:omega}
\end{figure}

We fit the radial profiles of the inverse screening length $\kappa$, temperature $T$, and plasma mass $\omega_{\rm p}$ in the cores of our MESA-generated stars with the following exponential functions
\begin{align}
    T&=T_c(M)e^{-\frac{r}{r_{T}(M)}},\label{T}\\
    \kappa&=\kappa_c(M)e^{-\frac{r}{r_{\kappa}(M)}},\label{kappa} \\
     \omega_{\rm p}&=\omega_{\text{p},c}(M)e^{-\frac{r}{r_{\omega_{\rm p}}(M)}}.\label{omega}
\end{align}
As displayed in Figs.~\ref{fig:T_fit}, \ref{fig:kappa_fit}, and \ref{fig:omega}, these exponential fits closely track the core profiles of the stars (where virtually all the axions are produced) but they start to fail near the surface of the stars. The stellar mass $M$ dependence of these parameters $T_c$, $r_T$, $\kappa_c$, $r_\kappa$, $\omega_{\text{p},c}$, $r_{\omega_p}$ are shown in Fig.~\ref{fig:parameters_fit}. We further fit these parameters as power laws in $M$
\begin{align}
    T_c(M)&=(1.83\pm0.06)  \left(\frac{M}{M_\odot}\right)^{0.22\pm0.01}\ \text{keV},\label{Tc}\\
    r_T(M)&=(0.86\pm0.03) \left(\frac{M}{M_\odot}\right)^{0.61\pm0.01}  R_\odot,\label{rT}\\
    \kappa_c(M)&= (10.4\pm 0.1)\left(\frac{M}{M_\odot}\right)^{-0.76\pm0.02}\ \text{keV},\label{kappac}\\
    r_\kappa(M)&=(1.06\pm0.05) \left(\frac{M}{M_\odot}\right)^{0.56\pm0.01}\ R_\odot,\label{rkappa}\\
    \omega_{\text{p},c}(M)&=(0.350\pm0.004)\left(\frac{M}{M_\odot}\right)^{-0.60\pm0.01}\ \text{keV},\label{wc}\\
    r_{\omega_{\rm p}}(M)&=(0.79\pm0.04) \left(\frac{M}{M_\odot}\right)^{0.57\pm0.01}  R_\odot,\label{rw}
\end{align}
where $M_{\odot}$ and $R_\odot$ are the mass and radius of the Sun. We also fit the main sequence lifetimes of our MESA stars as follows
\begin{align}
    t_{\rm life}(M)=\begin{cases}
        (6.8\pm0.1)\times10^{9}\left(\frac{M}{M_\odot}\right)^{-2.79\pm0.02}\ \text{yr},&M\lesssim10M_\odot\\
        (5\pm1)\times10^{7}\left(\frac{M}{M_\odot}\right)^{-0.63\pm0.08}\ \text{yr},&M\gtrsim10M_\odot
    \end{cases}\label{tlife}.
\end{align}
All the above stellar mass scalings are accurate only for massive stars with masses $1-100 M_\odot$ (which dominate the axion sourcing). Lower mass stars behave differently for at least a couple of reasons. Their nuclear burning is dominated by the p-p chain reaction instead of the CNO cycle \cite{Salaris:2005cno}. Moreover, they evolve very slowly and consequently fail to arrive at the intermediate main sequence age (the point where the hydrogen abundance is $X=0.3$) within the age of the universe. As mentioned in the main text, in such cases we extract the stellar profiles at half the age universe instead. 

\begin{figure}
    \centering
    \vspace{-1cm}
    \includegraphics[width=\linewidth]{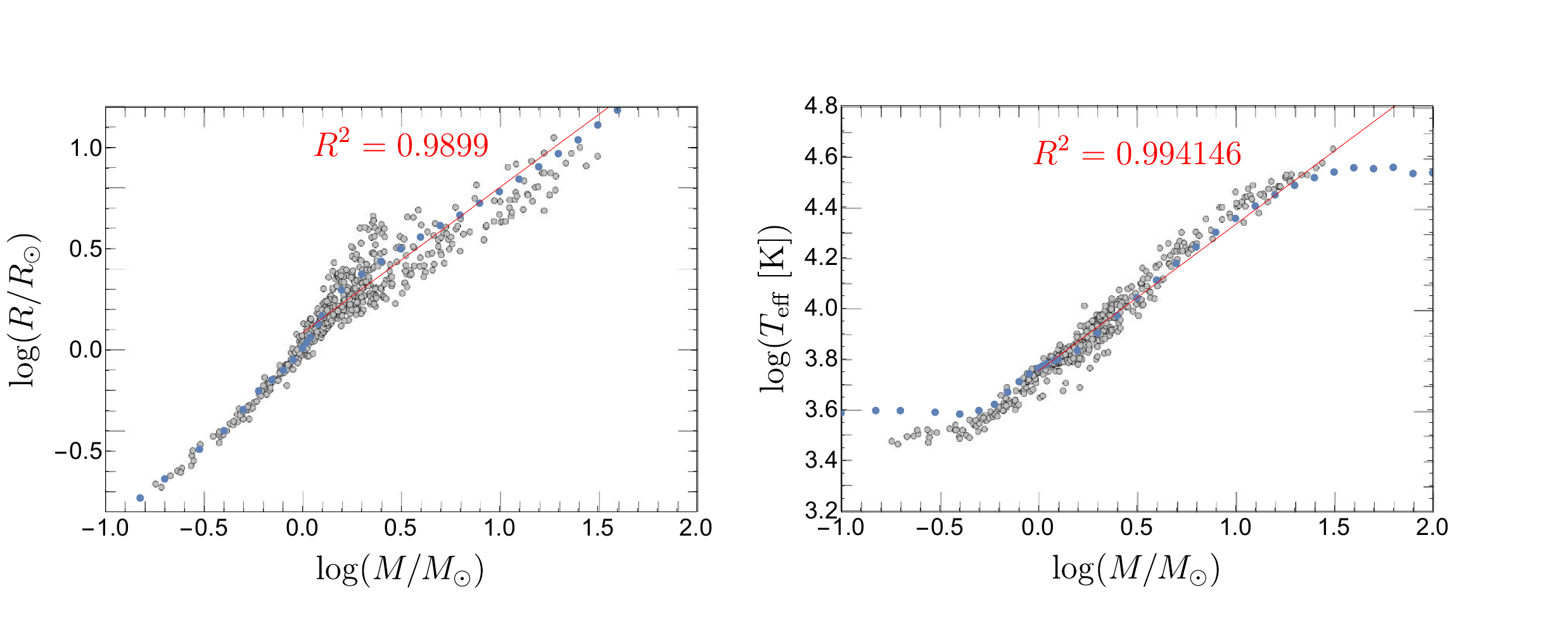}\vspace{-0.2cm}
    \caption{Visualisations of the fits in Eqs.\eqref{R*} and \eqref{Teff}. Blue dots are the data from MESA while the red lines are their power law fits. Gray dots are the observed data from \cite{Eker:2018}.}
       \label{fig:RTfit}
\end{figure}

To verify the accuracy of our MESA simulations, we extract the stellar radii $R$ and effective (surface) temperatures $T_{\rm eff}$ of our MESA stars and fit them as power laws
\begin{align}
    R(M) &= (0.80\pm0.08) \left(\frac{M}{M_\odot}\right)^{0.83\pm0.03}\ R_\odot,\label{R*} \\
    T_{\rm eff}(M) &= (6100\pm300) \left(\frac{M}{M_\odot}\right)^{0.54\pm0.02}\ \text{K}.\label{Teff}    
\end{align}
We stress that these fits come purely from MESA simulations without any input from the observed luminosity - mass relation. We compare these observable quantities with the existing data from \cite{Eker:2018}, including 509 main sequence stars selected from the ``Catalog of Stellar Parameters from the Detached Double-Lined Eclipsing Binaries in the Milky Way'' by \cite{2015AJ....149..131E}, and found, as shown in Fig.~\ref{fig:RTfit}, that they agree well.

\begin{figure}
    \centering
    \includegraphics[width=0.95\linewidth]{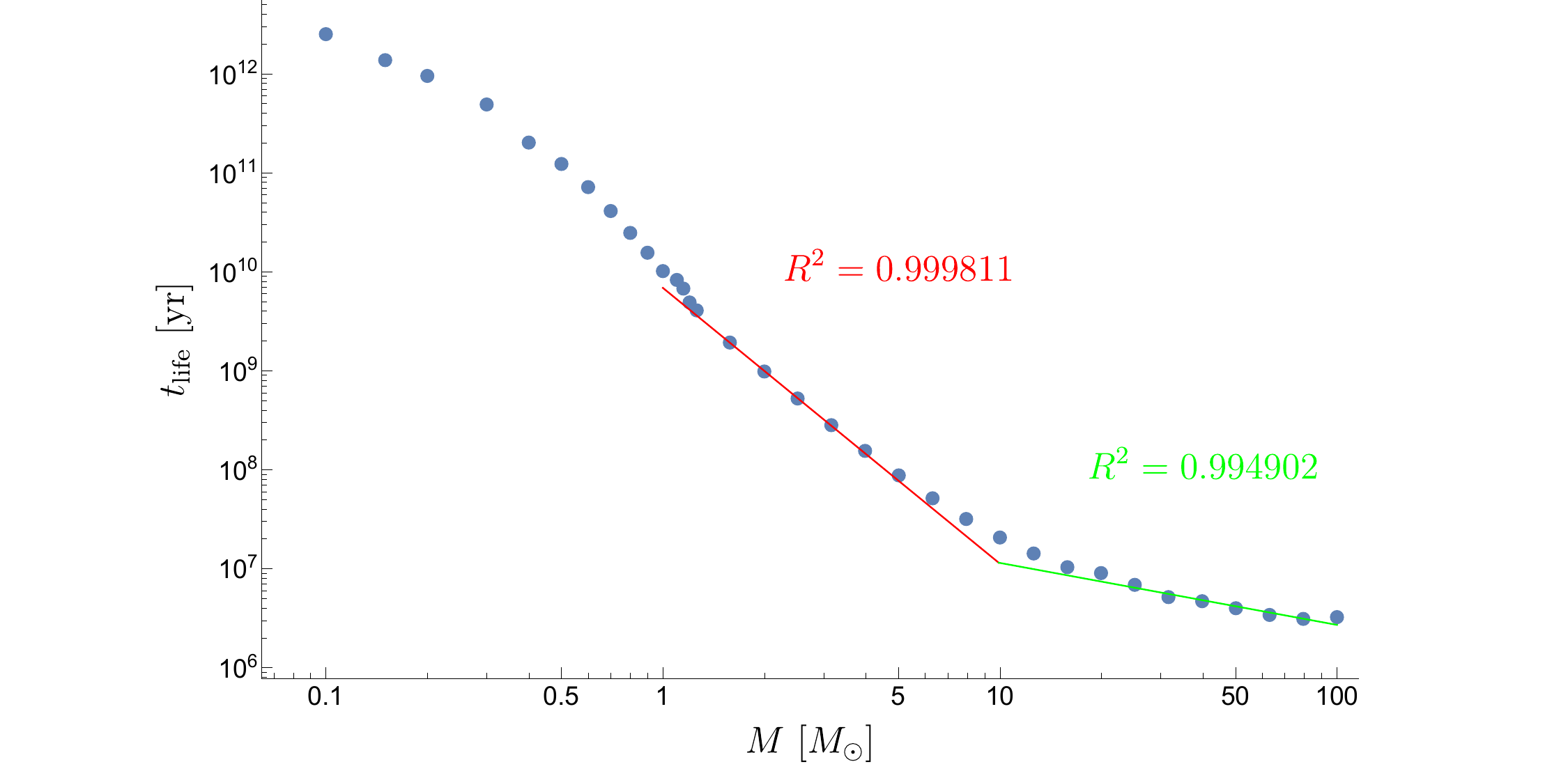}
    \caption{Visualisations of the fits in Eq.\eqref{tlife}. Blue dots are the data from MESA while the red and green lines are their broken (at $M\sim 10M_\odot$) power law fits.}
\end{figure}


\bibliography{references}
\bibliographystyle{JHEP}

\end{document}